\documentclass[a4paper,12pt]{article}
\pdfoutput=1

\usepackage{epsfig,graphicx,xcolor,amsbsy,amssymb,latexsym,amsfonts,amsmath,,tcolorbox,setspace}
\usepackage{pstricks}
\usepackage{color}
\usepackage{soul}
\usepackage[numbers,square,comma, compress]{natbib}
\usepackage{placeins}
\usepackage{wrapfig,framed,caption}
\usepackage[makeroom]{cancel}

\usepackage{wasysym}

\usepackage{multirow}

\definecolor{shadecolor}{gray}{0.925}

\usepackage{eurosym}
\usepackage[a4paper]{geometry}
\usepackage{hyperref}
\usepackage{graphicx}
\usepackage{tikz}
\usetikzlibrary{decorations.pathreplacing,decorations.markings,snakes}
\usetikzlibrary{decorations.pathmorphing}
\usetikzlibrary{patterns}
\usetikzlibrary{calc}
\usepackage{xcolor}
\usepackage{amsmath,amssymb,amsfonts,pstricks,setspace}
\usepackage[enableskew]{youngtab}

\numberwithin{equation}{section}

\newcommand{\bea}{\begin{eqnarray}\displaystyle}
\newcommand{\eea}{\end{eqnarray}}

\newcommand{\Icn}[1]{I_{\text{c},#1}}
\newcommand{\ms}{S}
\newcommand{\mi}{I}
\newcommand{\mr}{R}

\newcommand{\rin}{\gamma}
\newcommand{\rhe}{\epsilon}

\newcommand{\sig}{\sigma}

\newcommand{\op}[1]{\mathcal{O}_{#1}}

\newcommand{\pot}{\Phi}

\definecolor{cerulean}{rgb}{0.0, 0.48, 0.65}

\title{
\begin{flushright}{\vspace{-2.5cm}\small LYCEN 2021-01\\}\end{flushright}
\vspace{2.3cm}
{\bf XXX}\\[40pt]}

\author{\large Giacomo~Cacciapaglia$^{1,2}$\footnote{\tt g.cacciapaglia@ipnl.in2p3.fr},\; Corentin~Cot$^{1,2}$\footnote{\tt cot@ipnl.in2p3.fr},\; Adele de Hoffer$^{3}$\footnote{\tt s277940@studenti.polito.it},\\ 
Stefan~Hohenegger$^{1,2}$\footnote{\tt s.hohenegger@ipnl.in2p3.fr},\; Francesco Sannino$^{4,5,6}$\footnote{\tt sannino@cp3.sdu.dk}\; and Shahram  Vatani$^{1,2}$\footnote{\tt shahram.vatani@hotmail.fr}}

\title{
\begin{flushright}{\vspace{-2.5cm}\small LYCEN 2021-01\\}\end{flushright}
\vspace{2.3cm}
{\bf  { Epidemiological theory of virus variants}}\\[40pt]}

\begin{document}

\maketitle
\thispagestyle{empty}

\begin{center}
\renewcommand{\thefootnote}{\fnsymbol{footnote}}\vspace{-0.5cm}
${}^{1}$ Institut de Physique des 2 Infinis (IP2I) de Lyon, CNRS/IN2P3, UMR5822, \\ 69622 Villeurbanne, France\\[0.5cm]
\renewcommand{\thefootnote}{\fnsymbol{footnote}}\vspace{-0.5cm}
${}^{2}$ Universit\' e de Lyon, Universit\' e Claude Bernard Lyon 1, 69001 Lyon, France\\[0.5cm]
\renewcommand{\thefootnote}{\fnsymbol{footnote}}\vspace{-0.5cm} 
${}^{3}$  Politecnico di Torino, Torino, Italy  \\[0.5cm]
\renewcommand{\thefootnote}{\fnsymbol{footnote}}\vspace{-0.5cm} 
${}^{4}$  Scuola Superiore Meridionale, Largo S. Marcellino, 10, 80138 Napoli NA, Italy  \\[0.5cm]
\renewcommand{\thefootnote}{\fnsymbol{footnote}}\vspace{-0.5cm}
${}^{5}$ Dipartimento di Fisica, E. Pancini, Univ. di Napoli, Federico II and INFN sezione di Napoli, \\  Complesso Universitario di Monte S. Angelo Edificio 6, via Cintia, 80126 Napoli, Italy 
\renewcommand{\thefootnote}{\fnsymbol{footnote}}\vspace{-0.5cm} \\[0.5cm]
${}^{6}$ CP$^3$-Origins and D-IAS, Univ. of Southern Denmark,  Campusvej 55, DK-5230 Odense, Denmark
 \end{center}

\begin{center}
{\bf Abstract:} 
\end{center}
We propose a physical theory underlying the temporal evolution of competing virus variants that relies on the existence of (quasi) fixed points capturing the large time scale invariance of the dynamics. To motivate our result we first modify the time-honoured  compartmental models of the SIR type to account for the existence of competing variants and then show how their evolution can be naturally re-phrased in terms of flow equations ending at quasi fixed points. As the natural next step we employ (near) scale invariance to organise the time evolution of the competing variants within the effective description of the \emph{epidemic Renormalization Group} framework. We test the resulting theory against the time evolution of COVID-19 virus variants that validate the theory empirically.

\pagebreak
\tableofcontents

\section{Introduction}
In the wake of the ongoing pandemic of SARS-CoV-2, epidemiologists are currently witnessing a surge of data showing not only the spread and time evolution of the virus but also its genetic evolution, notably the emergence of mutations. While understanding their biological properties and assessing the danger they pose for humans is of tantamount importance, the very occurrence of new variants is also a crucial component in the time evolution of the pandemic itself. Indeed, the companion paper \cite{Companion} analyses the emergence of new variants as one of the driving forces behind the dynamics of the SARS-CoV-2 pandemic (notably the multi-wave structure) through the analysis of the above mentioned data by means of machine learning and numerical techniques. The current paper is aimed at providing a theoretical framework to model the interplay and competition of different virus variants within a given population in an efficient manner.

The application of mathematical modelling to describe and predict the spread of infectious diseases has a history going back more than a century \cite{Hamer,HamerLect1,HamerLect2,HamerLect3,Ross1911,Ross1916,RossHudson1916II,RossHudson1916III,McKendrick1912,McKendrick1914,McKendrick1926}. The pioneering SIR model \cite{Kermack:1927} is an example of compartmental models, which are based on dividing the population into different classes and model the spread of the disease via a set of first order differential equations in time that describe the flow of individuals between these different compartments. Such models are deterministic in the sense that the solutions are uniquely determined through the initial conditions supplemented to the differential equations (apart from certain parameters related to the infectivity of the disease and the rate at which individuals may recover from an infection). These models can be refined by further subdividing the population into more compartments depending on biological, geographical and social particularities of the situation. We refer the  reader to some of the excellent reviews that are available in the literature for more details, \emph{e.g.} \cite{PERC20171,WANG20151,WANG20161,HETHCOTErev,BaileyBook}.

Complementary to this, there exist other models of a stochastic nature: in these models, the microscopic processes leading to the spread of the disease are understood in a probabilistic sense and time is typically a discretised variable. Models of this type include lattice and percolation models, many of which are inspired by chemical or diffusion processes. We refer the reader to the excellent reviews \cite{Essam,Stauffer} for more details. These models are related to the compartmental models mentioned above through processes that effectively reduce the number of degrees of freedom, such as mean field approximations and averaging procedures.

While very different in their original modelling of the problem, the approaches outlined above, exhibit very interesting properties, such as criticality and symmetries related to a re-scaling of time. The former is the observation that in many solutions the difference between solutions where only a very small number of individuals is concerned compared to those where a significant fraction of the population becomes infected depends on certain threshold parameters and the transition is rather sharp. This similarity to phase transitions in physics has lead to further approaches using universality classes of field theories, such as in the pioneering works \cite{Domb,Peliti,Doi1,Doi2,Cardy_1985,Grassberger1983}.

Furthermore, a more effective incorporation of large and short time-scale invariance has been proposed in \cite{DellaMorte:2020wlc}. Here, the organisation of epidemic curves around temporal symmetry principles was termed the epidemic Renormalisation Group (eRG) framework  \cite{DellaMorte:2020wlc}. It stems from the formal identification of the running of the coupling strength in theories of fundamental interactions \cite{Wilson:1971bg,Wilson:1971dh} with the time dependence of epidemiological quantities, such as the cumulative number of infected \cite{DellaMorte:2020wlc}. After demonstrating the power of reorganising the epidemic diffusion process around time-scale invariances, the approach has been extended to account for human interactions and mobility across different regions of the world in \cite{Cacciapaglia:2020mjf}. When combined with mobility data provided by Google and Apple \cite{cacciapaglia2020mining} as well as US flight data \cite{cacciapaglia2020us}, the framework was used to deduce the impact of lockdowns on the global spread of the virus. This lead to the prediction, with few months of advance, of the advent of a second pandemic wave that started in the fall 2020 in Europe \cite{cacciapaglia2020second}. The framework has been extended to contain quasi-fixed points \cite{cacciapaglia2020evidence} to provide a first fully consistent mathematical description of multiwave pandemics. The extended framework, dubbed Complex eRG, related the interwave period to the timing of the insurgence of the next wave \cite{cacciapaglia2020multiwave}. Last but not least, a slight modification of the approach has shown effective to incorporate the first impact of the US vaccination campaign \cite{cacciapaglia2020us}.  

In the  comprehensive work of \cite{Cacciapaglia:2021vvu} we further showed how the eRG framework is related to  the traditional compartmental models of SIR type \cite{DellaMorte:2020qry}, and how they emerge from microscopic percolation models which are stochastic in nature.

The investigations above were purely epidemiological in nature, however with the advent of massive genome sequencing a new era commenced in which, as we shall see, these data become integral part of refined epidemiological models of the type discussed above. 

In order to acquire intuition on the virus variant diffusion and dynamics, in Section~\ref{Compartmental} we will use modified SIR-inspired compartmental models \cite{Kermack:1927}. In particular we will describe the competition between two virus variants in terms of a SIIR  model. Focussing on the temporal flow of the cumulative numbers of infected, the analysis  points to a more efficient description in terms of an eRG system of equations. From a theoretical physics standpoint, we find amusing the appearance of a degeneracy in the system, specifically in terms of the asymptotic number of infected, which can be interpreted as the emergence of a marginal operator regulating the end-point of the flow. These features seem peculiar to the rather special systems and will grant a deeper understanding at a more microscopic level. It will be interesting to study similar features in other 'microscopic' systems \emph{e.g.} such as percolation models described in \cite{Cardy_1985,Grassberger1983}. We therefore study the virus mutation version of the eRG (MeRG) in Section~\ref{Sect:RG}. We show that the flow-equations can be efficiently described in terms of a gradient flow diagrammatically. The flow diagram reveals the existence of fixed points to be interpreted as initial and final stages of the cumulative number of infected individuals by different virus variants. We discover that virus mutations and their variant evolutions can be efficiently represented in terms of theoretical physics concepts such as critical surfaces and (quasi)-fixed points.  We also discover that virus mutations are at the heart of switching on relevant operators from the perspective of fixed points controlling the final stage of a single wave pandemic. These lend a natural and profound interpretation of the complex fixed points eRG (CeRG) \cite{cacciapaglia2020evidence,cacciapaglia2020multiwave}, which mathematically models multi wave pandemics. 

To further substantiate our findings, in Section~\ref{Sect:RealWorld} we use the developed theory to describe epidemiological data of COVID-19 for several regions of the world where variants of concern have first emerged. The data on the genome sequencing was extracted from the online repository GISAID (\href{https://www.gisaid.org/}{gisaid.org/}), the epidemiological data for the US states was taken from the NY times \href{https://github.com/nytimes/covid-19-data/blob/master/us-states.csv}{github} and the epidemiological data for the other countries on \href{https://ourworldindata.org/}{Ourworldindata}.
We first used the GISAID data to extract the percentage of representation of the variants among the sequences collected at each specific date. Then, by multiplying this percentage to the epidemiological data on the number of new cases within the country, we estimated the number of new cases related to each variant. This method allowed us to generate the cumulative number of cases in order to fit it with a logistic function and extract the associated infection rate.

\section{Compartmental Models as Mutation Templates}
\label{Compartmental}

The eRG framework and its potential generalisations to several pathogens (\emph{e.g.} virus variants or different bacteria types) are effective descriptions of the system, capturing its dynamics via certain key fields (typically the cumulative number of infected individuals within a population). As such, other (`microscopic') degrees of freedom of the system have been `integrated out' (see \cite{Cacciapaglia:2021vvu} for more details on this point of view). While this approach is very efficient in simulating infection numbers, it is non-trivial to model new effects of the dynamics (such as the emergence of variants via mutations). To this end, we find it convenient to first obtain intuition by studying more basic models, of a compartmental nature, before using the obtained knowledge to propose eRG-like models. 

\subsection{Simple SIIR Model}

As a very simple `first-principle' model, we consider a compartmental model to describe the temporal evolution of two variants of a disease within a population of size $N$. The latter is split into four different classes (compartments):
\begin{itemize}
\item \emph{Susceptible individuals:} we denote by $N\,\ms(t)$ the number of individuals who can become infected with either variant of the disease.
\item \emph{Infectious individuals:} we denote by $N\,\mi_1(t)$ and $N\,\mi_2(t)$ the number of individuals that are currently infected with the two variants of the disease, respectively. Individuals in these two compartments can infect susceptible individuals, with the same variant of the disease, if getting in contact with them, with a well-defined constant rate. We assume that it is not possible to be infected with both variants simultaneously.
\item \emph{Removed individuals:} we denote by $N\,\mr(t)$ the number of individuals that can neither be infected themselves with any of the two variants, nor infect susceptible individuals. While this removal may not only be due to recovery from a previous infection with either of the two invariants, we assume that the latter grants permanent immunity with respect to both variants.
\end{itemize}
We assume that $N$ is sufficiently large such that the \emph{relative} numbers of susceptible $\ms(t)$, infectious $\mi_{1,2}(t)$ and removed individuals $\mr(t)$ are continuously differentiable functions of time
\begin{align}
\ms\,,\mi_{1,2}\,,\mr:\,\mathbb{R}_+\longrightarrow [0,1]\,.
\end{align}
Without loss of generality, we place the outbreak of the epidemic at time $t=0$. Furthermore, we consider the population to be stable in time, \emph{i.e.} we impose the constraint
\begin{align}
&\ms(t)+\mi_1(t)+\mi_2(t)+\mr(t)=1\,,&&\forall t\in\mathbb{R}_+\,.
\end{align}
The dynamics of the epidemic is described by individuals being moved between the four compartments introduced above with certain fixed\footnote{\label{fnote1}As we shall explain below, this assumption is one of the reasons why this model is not very suited to describe realistic epidemics and confront real world data. However, it simplifies the analysis while still providing some intuition on the dynamics of two competing variants of a disease.} (\emph{i.e.} time-independent) rates:
\begin{itemize}
\item $\rin_{1,2}$ are the rates at which susceptible individuals turn to infected once in contact with an individual infected by the two variants of the disease respectively.
\item $\rhe_{1,2}$ are the removal rates at which infectious individuals (carrying variants 1 or 2 respectively) become non-infectious. This includes recovery from the disease as well as other removal mechanisms (\emph{e.g.} death of the individual). 
\end{itemize}

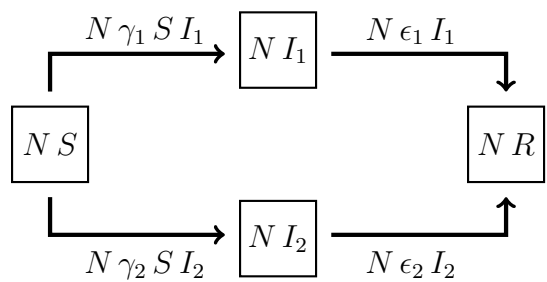
\begin{figure}[htbp]
\begin{center}
\scalebox{1}{\parbox{7cm}{\begin{tikzpicture}
\draw[thick] (0,0) rectangle (1,1);
\node at (0.5,0.5) {$N\,\ms$};
\begin{scope}[xshift=3cm,yshift=1.25cm]
\draw[thick] (0,0) rectangle (1,1);
\node at (0.5,0.5) {$N\,\mi_1$};
\end{scope}
\begin{scope}[xshift=3cm,yshift=-1.25cm]
\draw[thick] (0,0) rectangle (1,1);
\node at (0.5,0.5) {$N\,\mi_2$};
\end{scope}
\begin{scope}[xshift=6cm]
\draw[thick] (0,0) rectangle (1,1);
\node at (0.5,0.5) {$N\,R$};
\end{scope}
\draw[ultra thick,->]  (0.5,1.2) -- (0.5,1.7) -- (2.8,1.7);
\draw[ultra thick,->]  (4.2,1.7) -- (6.5,1.7) -- (6.5,1.2);
\draw[ultra thick,->]  (0.5,-0.2) -- (0.5,-0.7) -- (2.8,-0.7);
\draw[ultra thick,->]  (4.2,-0.7) -- (6.5,-0.7) -- (6.5,-0.2);
\node at (1.75,2) {$N\,\rin_1\, \ms\,\mi_1 $};
\node at (5.25,2) {$N\,\rhe_1\,\mi_1 $};
\node at (1.75,-1.1) {$N\,\rin_2\, \ms\,\mi_2 $};
\node at (5.25,-1.1) {$N\,\rhe_2\,\mi_2 $};
\end{tikzpicture}

}}
\caption{Schematic flow between the four compartments of the SIIR model.}
\label{Fig:FlowFigureSIIR}
\end{center}
\end{figure}

\noindent
Schematically,  the flow between the compartments $(\ms,\mi_1,\mi_2,\mr)$ is shown in Figure~\ref{Fig:FlowFigureSIIR}, and it can be 
described by the following system of coupled first order differential equations
\begin{align}
\frac{d\ms}{dt}(t)&=-(\rin_1\,\mi_1(t)+\rin_2\,\mi_2(t))\,\ms(t)\,,\nonumber\\
\frac{d\mi_1}{dt}(t)&=\rin_1\,\mi_1(t)\,\ms(t)-\rhe_1\,\mi_1(t)\,,\nonumber\\
\frac{d\mi_2}{dt}(t)&=\rin_2\,\mi_2(t)\,\ms(t)-\rhe_2\,\mi_2(t)\,,\nonumber\\
\frac{d\mr}{dt}(t)&=\rhe_1\,\mi_1(t)+\rhe_2\,\mi_2(t)\,,\label{DiffEqs4CompartNoRe}
\end{align}
together with the following initial conditions
\begin{align}
&\ms(0)=\ms_0\,,&&\mi_1(0)=\mi_{1,0}\,,&&\mi_{2}(0)=\mi_{2,0}\,,&&\mr(0)=0\,,&&\text{with}&&\begin{array}{l}0\leq \ms_0,\mi_{1,0},\mi_{2,0}\leq 1\,,\\[4pt] \ms_0+\mi_{1,0}+\mi_{2,0}=1\,.\end{array}\label{BoundaryConditionsSIIR}
\end{align}

\noindent
Concretely we shall consider variant 2 to have been created through a mutation of variant 1. 
To quantify the effect of the second variant on the evolution of the pandemic, we will study the evolution of the number of infected under the variant 1, $\mi_1 (t)$ for $\mi_{2,0} \neq 0$, and compare the results with the control case, defined in absence of the second variant, \emph{i.e.} $\mi_{2,0} = 0$. To better connect the results in this section with the eRG approach, we focus on the cumulative number of infected $\Icn{1}(t)$, defined as
\begin{align}
&\Icn{i}(t)=N\,\mi_{0,i}+N\,\rin_i\,\int_0^t\mi_{i}(t')\,\ms(t')\,dt'\,,&&\forall i=1,2\,.\label{DefCumulativeI}
\end{align}

We remark that the model introduced above does not trace the origin of the variant 2 of the virus as a mutation of the first one. The latter could be encoded in a more evolutionary version of the 4-compartment model, where a mutated variant could appear at some time $t_0>0$. To be concrete, the evolutionary model can be described by the following set of differential equations:
\begin{align} \label{eq:SIIRbeta}
&\frac{d\ms}{dt}=-(\rin_1\,\mi_1+\rin_2\,\mi_2)\,\ms\,,&&\text{with} &&\ms(0)=\ms_0\,,\mi_1(0)=\mi_{1,0}\,,\mi_{2}(0)=\mr(0)=0\nonumber\\
&\frac{d\mi_1}{dt}=\rin_1\,\mi_1\,\ms-\rhe_1\,\mi_1-\beta(t)\,\mi_1\,,&& &&0\leq \ms_0,\mi_{1,0},\leq 1\nonumber\\
&\frac{d\mi_2}{dt}=\rin_2\,\mi_2\,\ms-\rhe_2\,\mi_2+\beta(t)\,\mi_1\,,&& &&\ms_0+\mi_{1,0}=1\nonumber\\
&\frac{d\mr}{dt}=\rhe_1\,\mi_1+\rhe_2\,\mi_2\,,\end{align}
where $\beta(t)$ is a time-dependent rate with point-like support that converts at time $t_0>0$ (the instant in which the variant appears) a fraction of the infectious individuals of variant 1 into those of variant 2. This function is defined as:
\begin{align}
\beta(t)=\left\{\begin{array}{lcl}\beta_0 & \text{if} & t=t_0\,,\\ 0 &\text{if} & t\neq t_0\,.\end{array}\right.
\end{align}
However, since we are excluding the possibility of a re-infection of removed individuals (with either of the two variants) and we are only interested in the impact of the appearance of the mutation on the dynamics of variant 1 (\emph{i.e.} the evolution for $t>t_0$), the above SIIR model (\ref{eq:SIIRbeta}) is equivalent to the simpler system (\ref{DiffEqs4CompartNoRe}) for a suitable choice of the boundary conditions (\ref{BoundaryConditionsSIIR}).

\subsection{Impact on the Cumulative Number of Infected}\label{Sect:ImpactCumulInfect}

While an analytic solution of the system (\ref{DiffEqs4CompartNoRe}) and (\ref{BoundaryConditionsSIIR}) appears very difficult, we can study numerical solutions using a simple Euler integration method. Since to this end we have to consider fixed values for $\rin_{1,2}$ and $\rhe_{1,2}$, we first study the dependence of $\Icn{1}(t)$ on $\rin_{2}$ and $\rhe_2$. In analogy to the SIR model, it is convenient to define the reproduction numbers of the two variants as follows:
\begin{equation}
R_{0,1} \equiv \sigma_1 = \frac{\gamma_1}{\epsilon_1}\,, \qquad  R_{0,2} \equiv \sigma_2 = \frac{\gamma_2}{\epsilon_2}\,. 
\end{equation}
As we will see, the time dependence of the cumulative number of infected mainly depends on the two reproduction numbers instead of the detailed values of infection and recovery rates, under certain conditions.

\begin{figure}[htbp]
\begin{center}
\includegraphics[width=7.5cm]{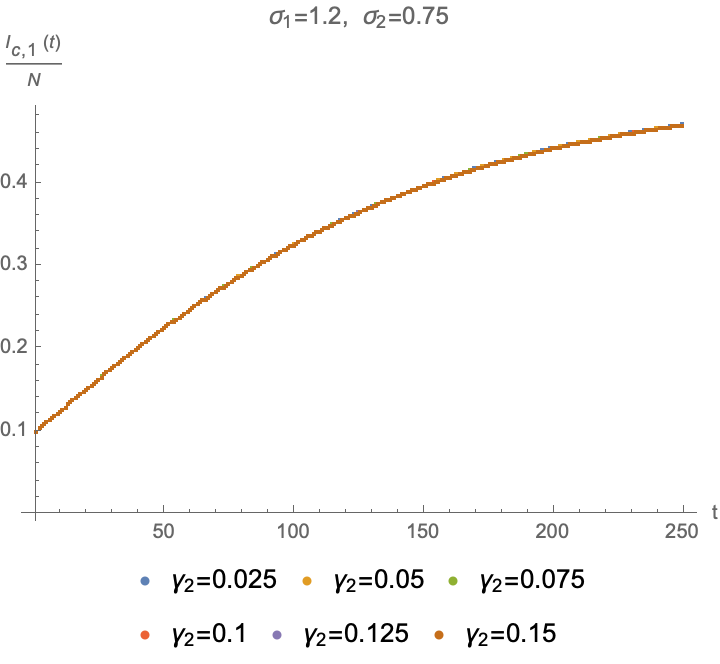}\hspace{0.5cm}\includegraphics[width=7.5cm]{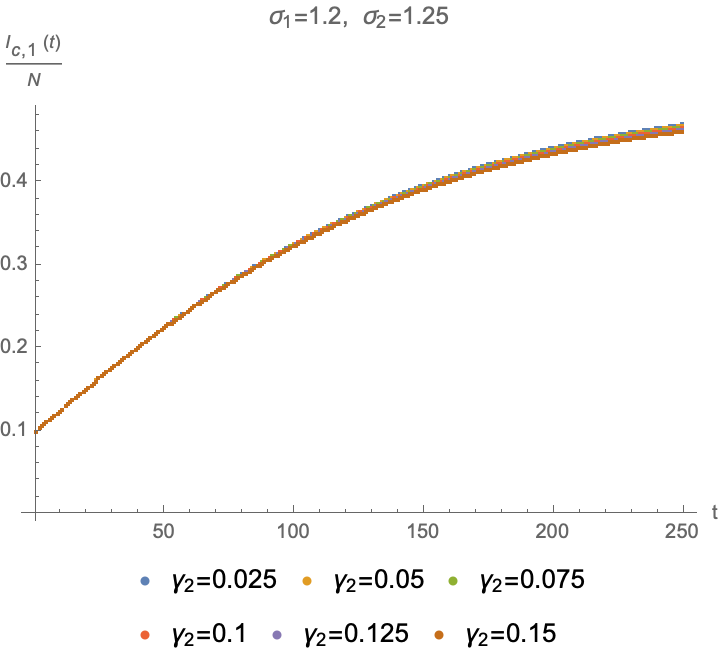}\\[0.5cm]\includegraphics[width=7.5cm]{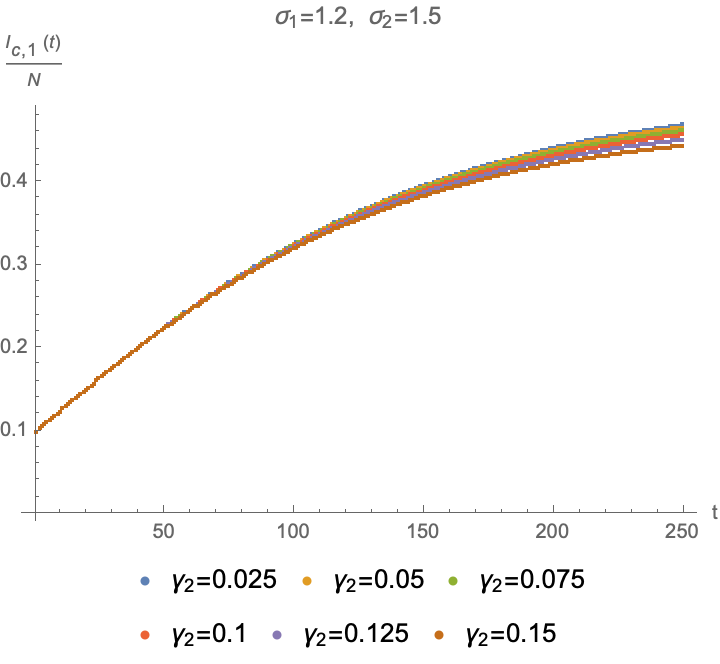}\hspace{0.5cm}\includegraphics[width=7.5cm]{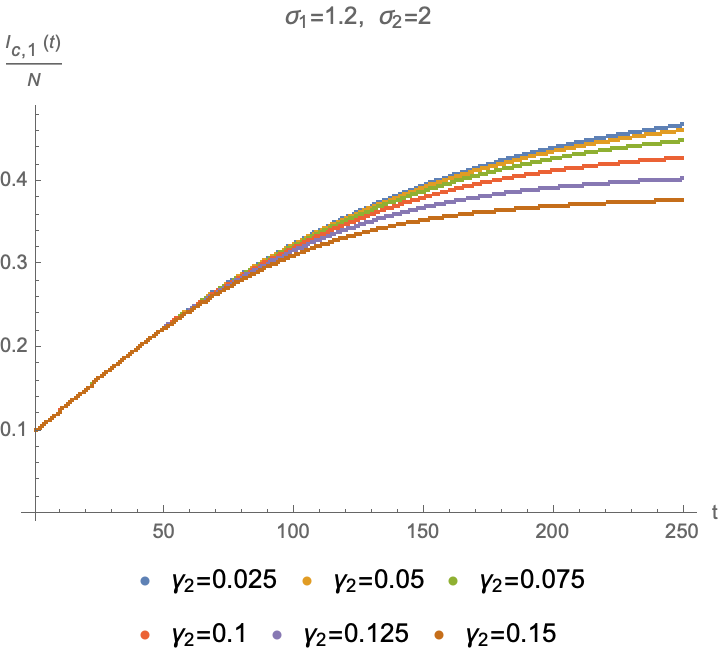}
\end{center}
\caption{Time evolution of $\Icn{1}$ as a function of $\rin_2$ with $\sig_2=0.75$ (top left panel), $\sig_2=1.25$ (top right panel), $\sig_2=1.5$ (bottom left panel)  and $\sig=2$ (bottom right panel). In all cases we have used $\rin_1=0.03$ and $\rhe_1=0.025$ (such that $\sig_1=1.2$ ) as well as $\ms_0=0.9$, $\mi_{1,0}=0.099$ and $\mi_{2,0}=0.001$.}
\label{Fig:VariantSigmaDependence}
\end{figure}

This point is illustrated in Figure~\ref{Fig:VariantSigmaDependence}, where we show solutions $\Icn{1}(t)$ for fixed $\sig_1 = 1.2$ and for four values of $\sigma_2$.  The various curves in each panel correspond to different values of $\gamma_2$ (where $\epsilon_2$ also varies accordingly to keep $\sigma_2$ fixed). These results, along with further checks we have performed, suggest that the cumulative number of infected by the original variant only depends on $\sigma_2$ as long as its value is similar to that of $\sigma_1$, i.e. if the two variants have similar reproduction numbers (quantitatively, $\sigma_2$ can be at most $\sim 50\%$ larger than $\sigma_1$). Conversely, for $\sigma_2 \gg \sigma_1$, the solutions of the SIIR model are sensitive to the specific values of $\gamma_2$ and $\epsilon_2$.
The results in Fig.~\ref{Fig:VariantSigmaDependence} also highlight that the shape of $\Icn{1}(t)$, or equivalently its asymptotic value at large times,  is substantially affected by the presence of the second variant only is $\sigma_2$ is significantly larger than $\sigma_1$.

\begin{figure}[htbp]
\begin{center}
\includegraphics[width=5.5cm]{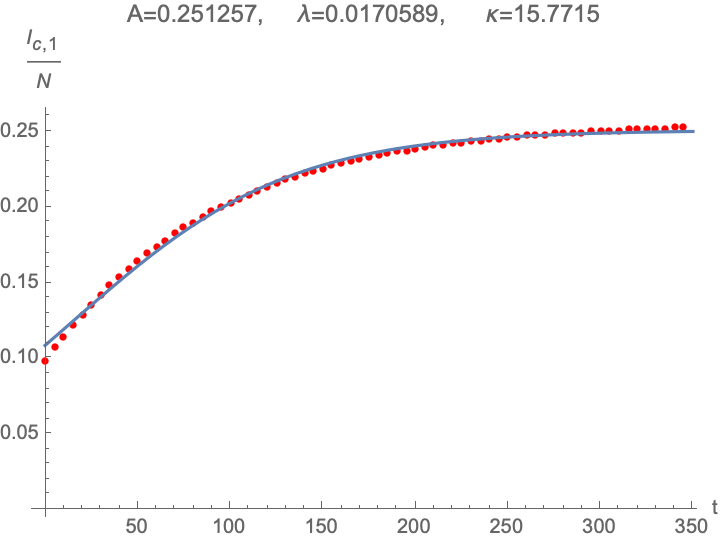}\hspace{0.1cm}\includegraphics[width=5.5cm]{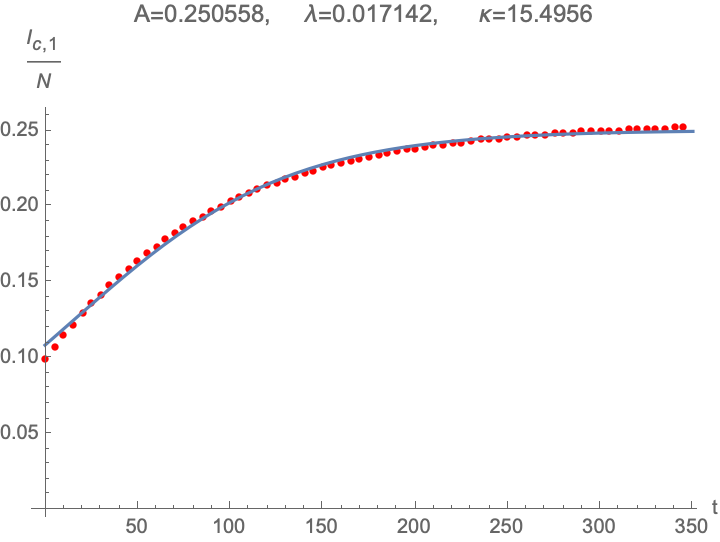}\hspace{0.1cm}\includegraphics[width=5.5cm]{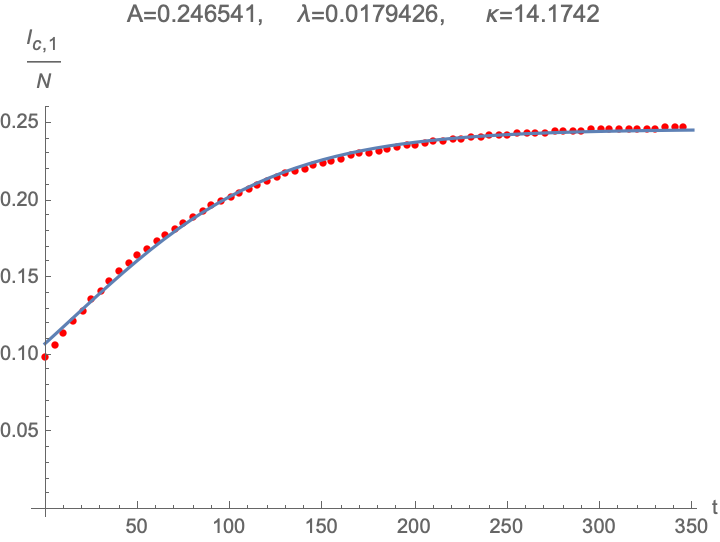}\\[0.5cm]\includegraphics[width=5.5cm]{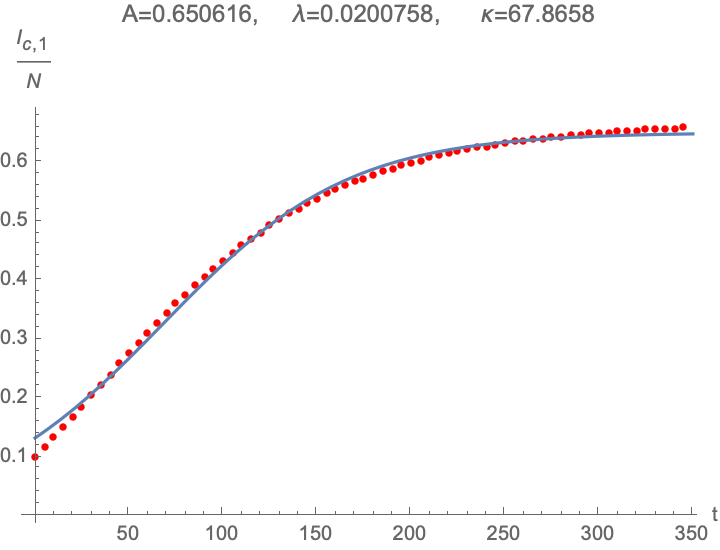}\hspace{0.1cm}\includegraphics[width=5.5cm]{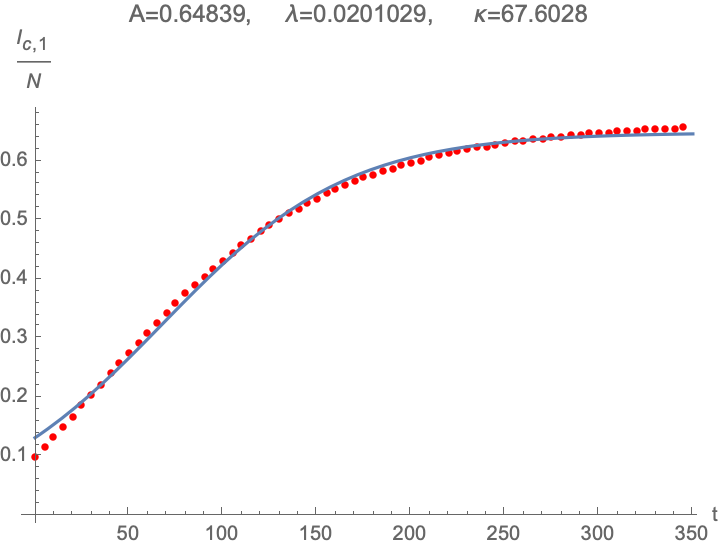}\hspace{0.1cm}\includegraphics[width=5.5cm]{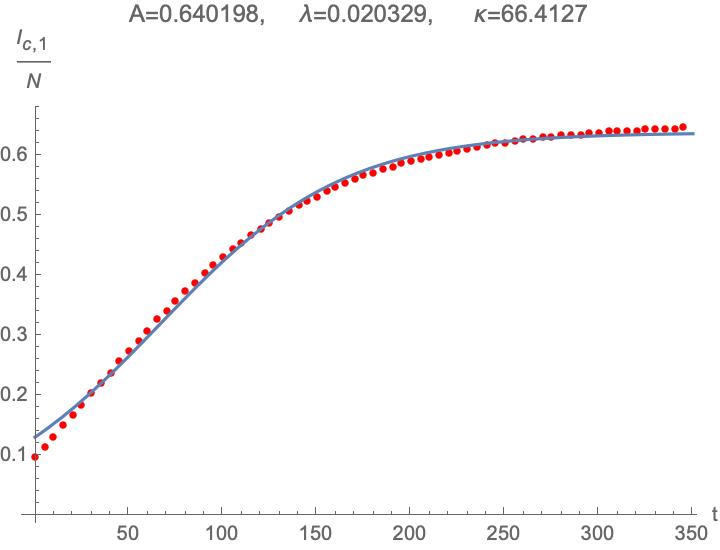}\\[0.5cm]
\includegraphics[width=5.5cm]{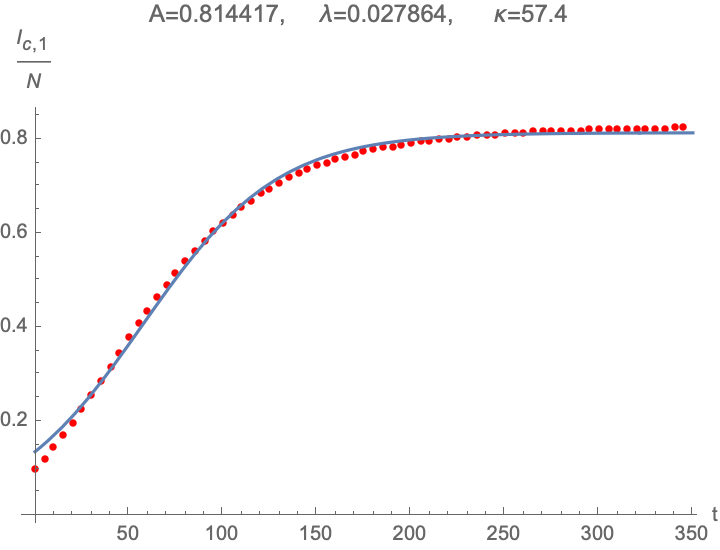}\hspace{0.1cm}\includegraphics[width=5.5cm]{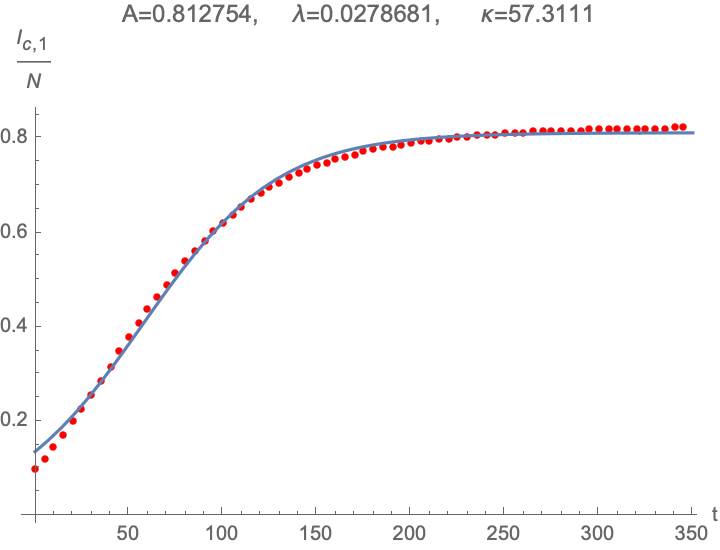}\hspace{0.1cm}\includegraphics[width=5.5cm]{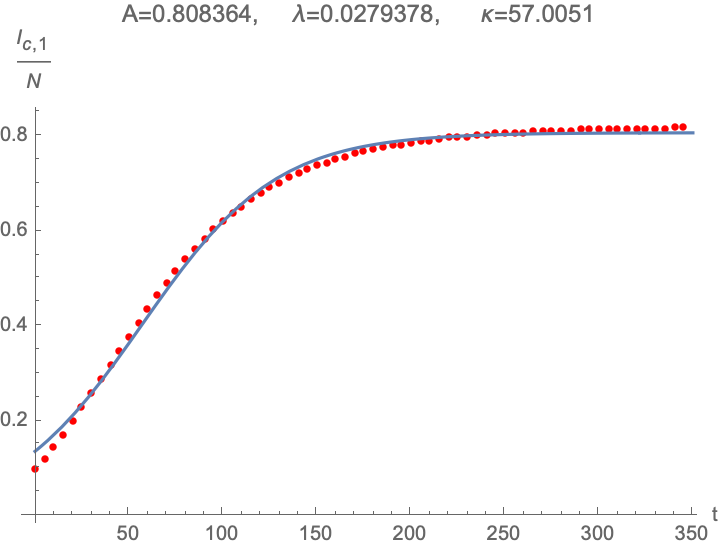}
\end{center}
\caption{Numerical solution (blue) and fitted logistic function (red) of $\Icn{1}$ as a function of time for two different combinations of $\sig_{1,2}$: top row: $(\sig_1,\sig_2)=(0.75,0.5)\,,(0.75,1.25)\,,(0.75,2)$, middle row: $(\sig_1,\sig_2)=(1.5,0.5)\,,(1.5,1.25)\,,(1.5,2)$, bottom row: $(\sig_1,\sig_2)=(2,0.5)\,,(2,1.25)\,,(2,2)$. All plots use $\ms_0=0.9$, $\mi_{1,0}=0.099$, $\mi_{2,0}=0.001$.}
\label{Fig:FitLogisticFunctions}
\end{figure}

To quantify the observations above, we can use the fact that the numerical solutions for $\Icn{1}(t)$ are fairly well approximated by a logistic function of the form
\begin{align}
&\Icn{1}(t)\sim \frac{N\,A}{1+e^{-\lambda (t-\kappa)}}\,,&&\text{with}&&\begin{array}{lcl}A\in[0,1]\,,\\[4pt]\lambda,\kappa\in\mathbb{R}\,,\end{array}\label{LogisticForm}
\end{align}
for different values of the parameters $(A,\lambda,\kappa)$. Here, $N\,A$ is the asymptotic value of the cumulative number of infected
\begin{align}
N\,A=\lim_{t\to\infty}\,\Icn{1}(t)\,,
\end{align}
while $\lambda$ is an effective infection rate, \emph{i.e.} a measure of how infectious the variant is (and is generally related to $\sig_1$), while $\kappa$ is a parameter that allows to shift the outbreak of the spread of variant 1. Exemplary fits of numerical solutions of $\Icn{1}(t)$ for different values of $(\sig_1,\sig_2)$ along with the numerical fitting parameters are shown in Figure~\ref{Fig:FitLogisticFunctions}. We remark, however, that fitting the cumulative number of infected individuals with logistic functions is not the only viable approximation and we shall encounter other possibilities (depending on specific cases) in Section~\ref{Sect:SIIRrg}.

\begin{figure}[h]
\begin{align}
&\parbox{5.5cm}{\epsfig{file=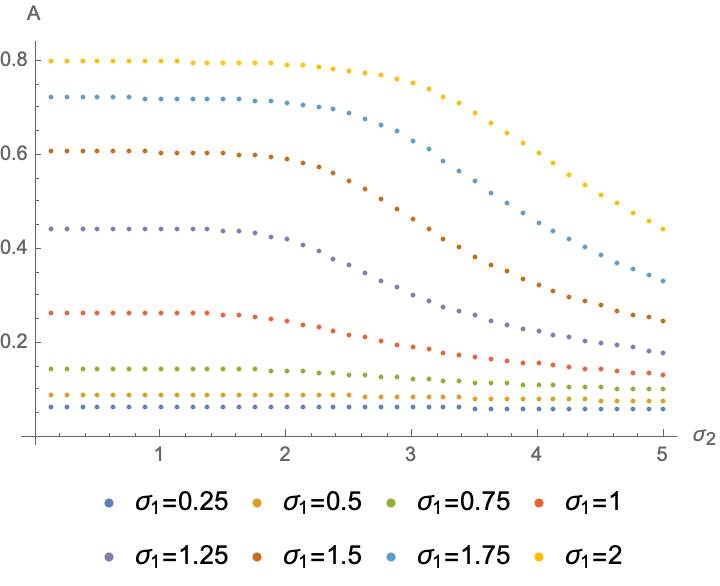,width=5.5cm}} && \parbox{5.5cm}{\epsfig{file=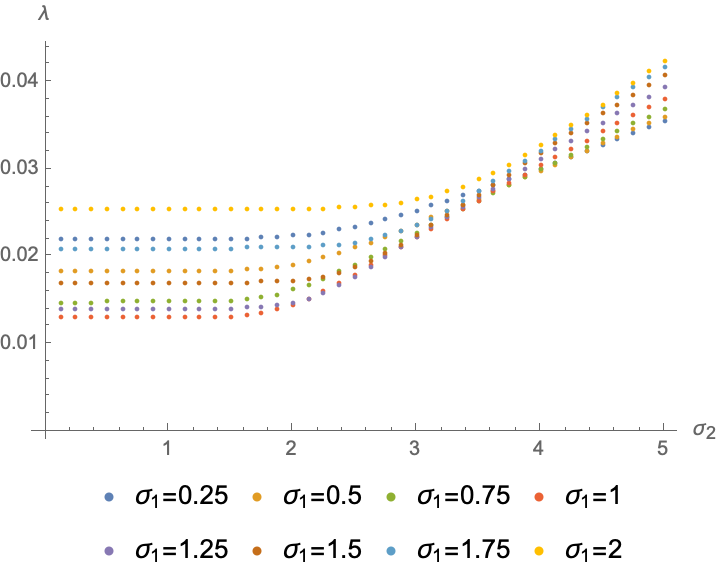,width=5.5cm}}&&\parbox{5.5cm}{\epsfig{file=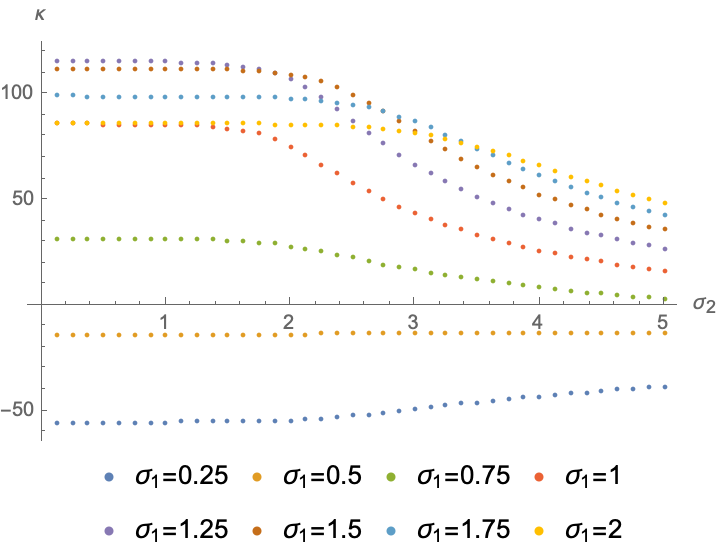,width=5.55cm}} \nonumber
\end{align}
\caption{Fit parameters $(A\,,\lambda\,,\kappa)$ for different values of $\sig_1$ as functions of $\sig_2$. All plots are shown for $\ms_0=0.9$, $\mi_{1,0}=0.099$, $\mi_{2,0}=0.001$.}
\label{Fig:ScanLogistic}
\end{figure}

Finally, using the result that $\Icn{1}$ can be approximated using a logistic function, we can quantify more concretely the impact of the appearance of variant 2. To this end, we plot in Figure~\ref{Fig:ScanLogistic} the values $(A,\lambda,\kappa)$ of the approximations of $\Icn{t}$ as functions of $\sig_2$. These examples suggest that an existing variant of a disease is strongly impacted by the appearance of a new variant only if the latter is significantly more infectious, \emph{i.e.} if $\sig_2$ is much larger than $\sig_1$ (in many cases approximately by a factor of 2).

We can summarise the basic findings suggested by the simple model (\ref{DiffEqs4CompartNoRe}) and (\ref{BoundaryConditionsSIIR})  through the following three points:

\begin{itemize}
\item Despite not being an exact solution, logistic functions (sigmoids) are good approximations to describe competing variants of a disease.
\item Key parameters of a variant in competition with a second variant that is not significantly more infectious can be described, to first approximation, by the reproduction numbers $\sigma_{1,2}$ alone (rather than individually $\rin_{1,2}$ and $\rhe_{1,2}$).
\item In order to have a significant impact on an existing variant (\emph{i.e.} to change key parameters such as the asymptotic numbers of infected individuals, the effective infection rates, \emph{etc}.), a new variant needs to have a much higher reproduction number, $\sigma_2 \gg \sigma_1$.
\end{itemize}

\subsection{Towards RG Flows}\label{Sect:SIIRrg}

\begin{figure}[htbp]
\begin{center}
\includegraphics[width=7.5cm]{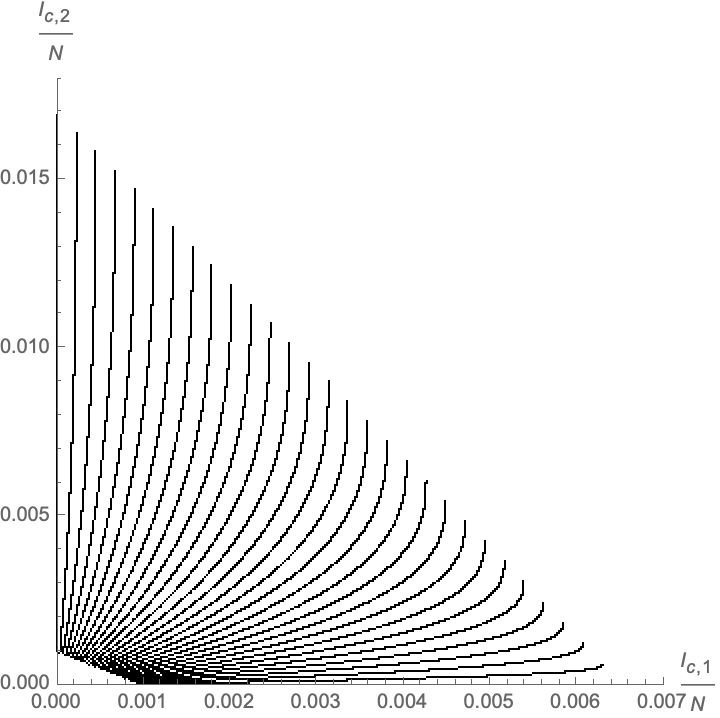}\hspace{1cm}\includegraphics[width=7.5cm]{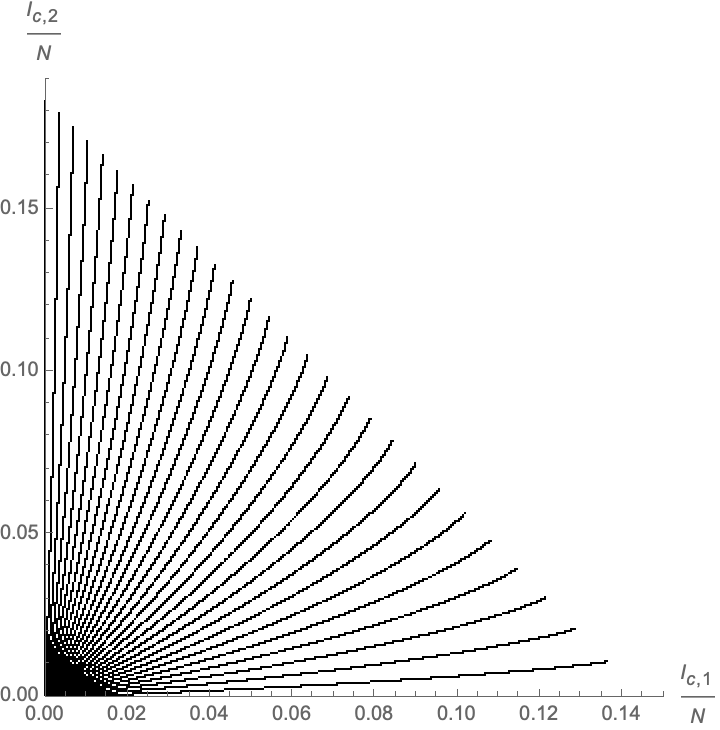}
\end{center}
\caption{Trajectories for different initial conditions in the $\mathbb{P}$ plane for $\ms_0=0.999$ and $(\sig_1,\sig_2)=(0.85,0.95)$ (left panel) and $(\sig_1,\sig_2)=(1.075,1.1)$ (right panel).}
\label{Fig:RGTrajectories}
\end{figure}

Having understood the impact of variants on the evolution of the infections in a simple SIIR model, we now want to understand the dynamics of the system in analogy to the renormalisation group in particle physics \cite{Wilson:1971bg,Wilson:1971dh}. This is a preliminary step that will help us gear up towards the inclusion of variants in the eRG framework. In practice, the solutions of the system of equations (\ref{DiffEqs4CompartNoRe}) and (\ref{BoundaryConditionsSIIR}) can be thought of as describing the flow of the physical system (here represented by the cumulative number of infected by the two variants) in the plane
\begin{align}
\mathbb{P}=\{(\Icn{1},\Icn{2})\in[0,N]\times[0,N]|\Icn{1}+\Icn{2}\leq N\}\,.
\end{align}
The trajectories are parameterised by time $t$. 
Following the eRG approach \cite{DellaMorte:2020wlc}, we shall try to interpret these trajectories as renormalisation group flows from a repulsive fixed point at $(\Icn{1},\Icn{2})=(0,0)$ to a new fixed point $(\Icn{1},\Icn{2})\neq (0,0)$ in the plane $\mathbb{P}$. In this regard, we shall consider the initial conditions (\ref{BoundaryConditionsSIIR}) as small deformations away from the initial fixed point. Keeping $\ms_0$ fixed, this gives a line (a co-dimension 1 surface) of starting points of trajectories in the plane $\mathbb{P}$, which can be found by solving (\ref{DiffEqs4CompartNoRe}) with the initial conditions satisfying 
\begin{align}
\ms_0=1-\mi_{1,0}-\mi_{2,0}\,.\label{CondsInit}
\end{align}
We then plot $(\Icn{1}(t),\Icn{2}(t))$ for successive times $t\in[0,\infty)$. The result is shown schematically, for different values of $\sig_{1,2}$, in Figure~\ref{Fig:RGTrajectories}. We remark that here (and in the following) we have assumed $\rhe_1=\rhe_2$, \emph{i.e.} that the removal rates for both variants are the same.\footnote{As we have seen in Section~\ref{Sect:ImpactCumulInfect}, choosing different $\rhe_{1,2}$ leads to only small deviations, as long as $\sig_1$ and $\sig_2$ are not too different from one another.}

Before analysing in more detail the cases of $\sig_{1,2}<1$ and $\sig_{1,2}>1$ (which show certain qualitative differences), we first remark that the flows, \emph{i.e.} the lines in Figure~\ref{Fig:RGTrajectories}, do not continue indefinitely but end at determined points in the $\mathbb{P}$-plane. These can be considered fixed points of the flow, as the evolution of the epidemic stops once these points are reached. More concretely, we observe that the line of initial conditions flows to another line of end-points. In fact, from the point of view of the SIIR model, the endpoints are equivalent, and correspond to the same final state of the system.
To understand this, we recall that the asymptotic solutions of the model (\ref{DiffEqs4CompartNoRe}) and (\ref{BoundaryConditionsSIIR}) corresponds to the eradication of both variants
\begin{align}
\lim_{t\to \infty}\mi_1(t)=0=\lim_{t\to\infty}\mi_2(t)\,.
\end{align}
Therefore at the endpoints in Figure~\ref{Fig:RGTrajectories} (which correspond to $t\to\infty$), only susceptible and removed individuals remain in the system and
\begin{align}
\lim_{t\to \infty} (\Icn{1}+\Icn{2})(t)=\lim_{t\to \infty} N\,R(t)\,.
\end{align}
Hence, the co-dimension 1 line of end-points in Figure~\ref{Fig:RGTrajectories} reflects different weighted re-distributions of the removed individuals into $\Icn{1}$ and $\Icn{2}$. In the SIIR model, however, the removed individuals are indistinguishable as they are collected in a single compartment $R$, and all the final configurations in the end-point line correspond to the same type of final state of the SIIR compartments (\emph{i.e.} with $\mi_1=\mi_2=0$). 
In other words, while the flow lines keep track of which individuals have been infected with which variant, this distinction becomes irrelevant at the level of the SIIR system when all previously infected individuals have become (indistinguishable) removed individuals. We shall see that this redundancy of the endpoints of the 'flow' of the system has a very natural interpretation from the perspective of the epidemic renormalisation group.

\subsubsection{Flow for $\sig_{1,2}<1$}
We shall now analyse in more detail the trajectories of the system in the $\mathbb{P}$-plane and we first focus on the case 
\begin{align}
&\sig_1:=\frac{\rin_1}{\rhe_1}<1\,,&&\text{and} &&\sig_2:=\frac{\rin_2}{\rhe_2}<1\,.\label{SigsSmaller}
\end{align}
In the simple SIR model \cite{Kermack:1927}, the values of $\sig_{1,2}$ correspond to the basic reproduction numbers of each variant. Physically, they tell the average number of (susceptible) individuals who are infected by a single infectious one during the period in which the latter is infectious. The condition $\sig_i<1$, therefore, implies that the number of infectious individuals $\mi_{i}(t)$ is a monotonically decreasing function of time, since $\frac{d\mi_{i}}{dt}(t)<0$ $\forall t>0$, as can be seen from eq.~(\ref{DiffEqs4CompartNoRe}).
 
\begin{figure}[htbp]
\begin{center}
\includegraphics[width=7.5cm]{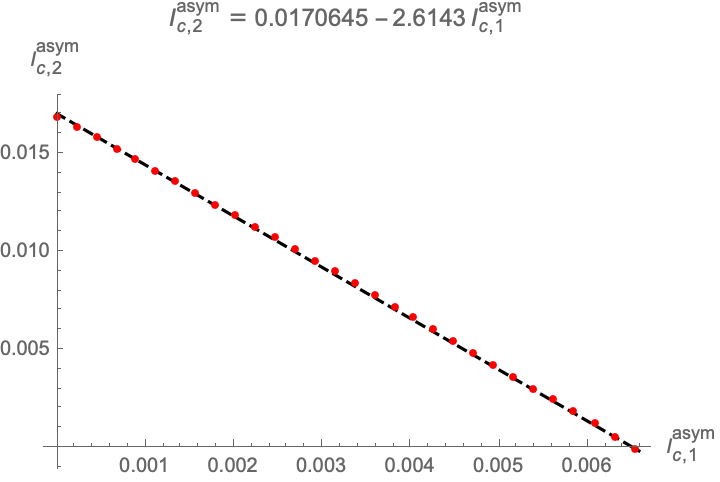}\hspace{1cm} \includegraphics[width=7.5cm]{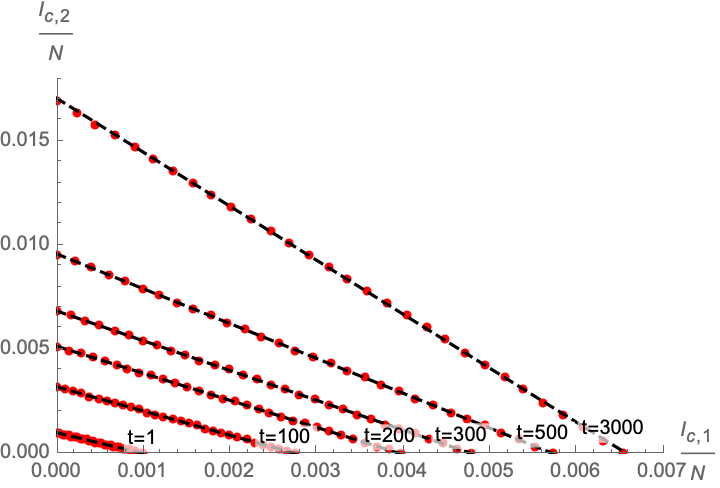}
\end{center}
\caption{Left panel: Approximation (\ref{AsymptoticLinear}) of the codimension 1 surface (dashed black line) that represents the end-points of the trajectories in the $\mathbb{P}$-plane. Each red dot represents $(\Icn{1}^{\text{asym}},\Icn{2}^{\text{asym}})$  at asymptotic time $t\to\infty$ for different choices of $(\mi_{1,0},\mi_{2,0})$ satisfying (\ref{CondsInit}). Right panel: Numerical approximations (dashed black lines) for equal time-slices $(\Icn{1}(t),\Icn{2}(t))$ for finite $t$ (red dots). Both plots use $\ms_0=0.999$ and $(\sig_1,\sig_2)=(0.85,0.95)$. }
\label{Fig:RGTrajectoriesAsymptoticInf}
\end{figure}

As is visible in the left panel of Figure~\ref{Fig:RGTrajectories}, under the condition (\ref{SigsSmaller}) the end points of the trajectories of the system follow an approximate linear relation in the $\mathbb{P}$-plane. Concretely, upon defining the positions of the end-points as
\begin{align}
&\Icn{i}^{\text{asym}}(\mi_{1,0},\mi_{2,0})=\frac{1}{N}\,\lim_{t\to\infty}\Icn{i}(t)\,,&&\forall i=1,2\,,\label{AsymptoticLimitIc}
\end{align}
which depend on the initial conditions $(\ms_0,\mi_{1,0},\mi_{2,0})$ subject to (\ref{CondsInit}), we find to good approximation (see Figure~\ref{Fig:RGTrajectoriesAsymptoticInf} for an example) the relation 
\begin{align}
&\Icn{2}^{\text{asym}}(\mi_{1,0},\mi_{2,0})\sim a\,\Icn{1}^{\text{asym}}(\mi_{1,0},\mi_{2,0})+b\,,&&\text{with} &&\begin{array}{l}a=-\frac{1+\frac{1}{\sig_2}\,W(-\ms_0 \sig_2\,e^{\sig_2})}{1+\frac{1}{\sig_1}\,W(-\ms_0 \sig_1\,e^{\sig_1})}\,,\\[10pt] b=1+\frac{1}{\sig_2}\,W(-\ms_0 \sig_2\,e^{\sig_2})\,,\end{array}\label{AsymptoticLinear}
\end{align}
where $W$ is the Lambert function. The coefficients $(a,b)$ can be inferred from the points $(\Icn{1}^{\text{asym}}(1-\ms_0,0),0)$ and $(0,\Icn{2}^{\text{asym}}(0,1-\ms_0))$, which can be determined analytically from the SIR model \cite{Kermack:1927} with 3 compartments (\emph{i.e.} with only one variant of the disease).  

In fact, even at finite time $t<\infty$, we have found numerical evidence for a linear relation between $\Icn{1}(t)$ and $\Icn{2}(t)$ with different initial conditions $(\mi_{1,0},\mi_{2,0})$: the right panel of Figure~\ref{Fig:RGTrajectoriesAsymptoticInf} shows equal time slices of the flows in the $\mathbb{P}$-plane, \emph{i.e.} $(\Icn{1}(t),\Icn{2}(t))$ with different $(\mi_{1,0},\mi_{2,0})$ evaluated at the same time $t<\infty$ (red points), which are approximated by linear curves (dashed black lines).

\begin{figure}[htbp]
\begin{center}
\includegraphics[width=7.5cm]{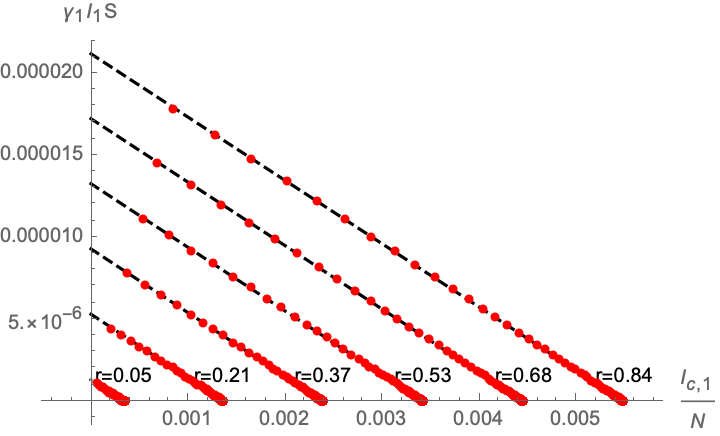}\hspace{1cm} \includegraphics[width=7.5cm]{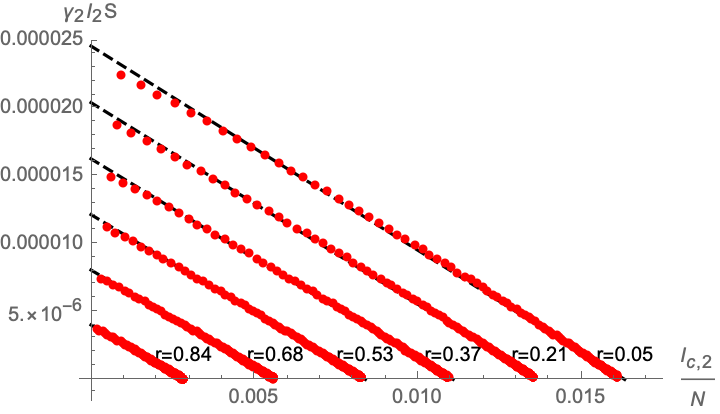}
\end{center}
\caption{Linear approximation of $(\Icn{1},\tfrac{d\Icn{1}}{dt})$ (left panel) and $(\Icn{2},\tfrac{d\Icn{2}}{dt})$ (right panel) as a function of the parameter $r$ (defined in eq.~(\ref{DefParamInit})). The red dots represent numerical solutions of $(\Icn{i},\tfrac{d\Icn{i}}{dt})$ for different values of time $t$, grouped according to the initial conditions. All plots use $\ms_0=0.999$ and $(\sig_1,\sig_2)=(0.85,0.95)$.}
\label{Fig:LinearBetaFunctionsSIIRInf}
\end{figure}

\begin{figure}[htbp]
\begin{center}
\includegraphics[width=10.5cm]{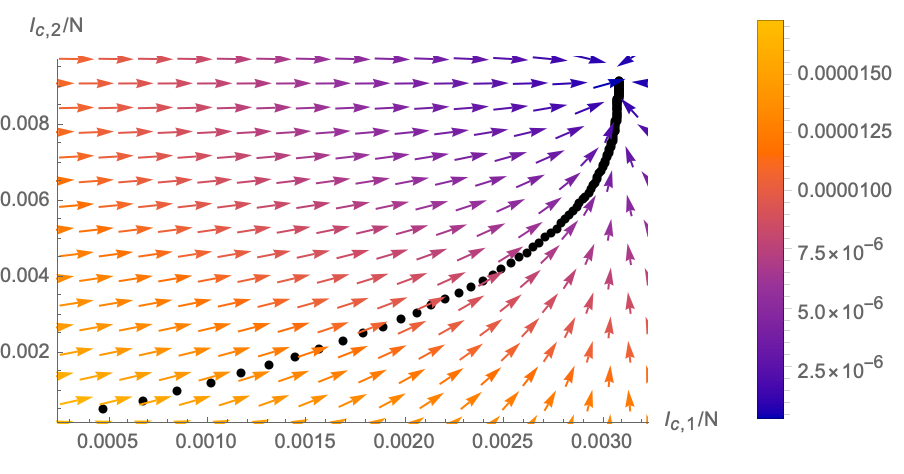}
\end{center}
\caption{Trajectory of the system in the $\mathbb{P}$-plane for a fixed initial condition (black line) following the vector field $\frac{1}{N}(\tfrac{d\Icn{1}}{dt},\tfrac{d\Icn{2}}{dt})$ (the colour of the vectors represents the length). We have used the numerical parameters $r=\tfrac{1}{2}$, $\ms_0=0.999$ and $(\sig_1,\sig_2)=(0.85,0.95)$.}
\label{Fig:RGTrajectoriesDiff}
\end{figure}

Next, in order to make contact with a possible eRG approach to describe the flow of the system in the $\mathbb{P}$-plane, we consider the time derivatives of $(\mi_{c,1}(t),\mi_{c,2}(t))$. According to the definition~(\ref{DefCumulativeI}) they are given by
\begin{align}
&\frac{d\Icn{i}}{dt}(t)=N\,\rin_i\,\mi_i(t)\,\ms_i(t)\,,&&\forall\,i=1,2\,.\label{DiffEqIcn}
\end{align}
After dividing by the total size of the population for simplicity, from studying the numerical solutions of (\ref{DiffEqs4CompartNoRe}) and (\ref{BoundaryConditionsSIIR}) we observe  that these time derivatives can be well approximated by linear functions in $\Icn{i}$
\begin{align}
&-\beta_i(\Icn{j}(t)):=\frac{1}{N}\,\frac{d\Icn{i}}{dt}(t)\sim\lambda_i\left(1-\frac{\Icn{i}(t)}{N\,A_i}\right)\,,&&\forall\,i=1,2\,,\label{BetaFunctionLinearDef}
\end{align}
where $(\lambda_i,A_i)$ are constants that are implicitly functions of the initial conditions $(\mi_{1,0},\mi_{2,0})$. Indeed, examples of the approximation are shown in Figure~\ref{Fig:LinearBetaFunctionsSIIRInf}. Here, in order to better gauge the impact of the initial conditions, we have parametrised them as
\begin{align}
&\mi_{1,0}=(1-\ms_0)\,r\,,&&\mi_{2,0}=(1-\ms_0)(1-r)\,,&&\text{with}\,&&r\in [0,1]\,.\label{DefParamInit}
\end{align}
In fact, one can also represent the trajectory of the system (for fixed initial conditions $(\mi_{1,0},\mi_{2,0})$) in the $\mathbb{P}$-plane as following the vector field $\frac{1}{N}(\tfrac{d\Icn{1}}{dt},\tfrac{d\Icn{2}}{dt})$, as is schematically shown in Figure~\ref{Fig:RGTrajectoriesDiff}. 
This plot better illustrates the concept of flow of the system in the two dimensional plane: the system slows down as it approaches the fixed point. This is represented by the colour code of the flow lines (arrows in Figure~\ref{Fig:RGTrajectoriesDiff}), matching the length of the derivative vector and is also visible from the fact that the black dots of the trajectory (which represent the system after equal time intervals) become denser as they approach the fixed point.

We can express the approximated derivatives (\ref{BetaFunctionLinearDef}) in the following form:
\begin{align}
&-\beta_i(\Icn{j})=\frac{1}{N}\,\frac{d\Icn{i}}{dt}(t)\sim\nabla_i\,\pot(\Icn{j}(t))\,,&&\text{with}&&\pot(\Icn{j})=\sum_{j=1}^2\Icn{j}\,\lambda_j\left(1-\frac{\Icn{j}}{2NA_j}\right)\label{BetaLin}
\end{align}
where we have used that $\beta_i$ in (\ref{BetaFunctionLinearDef}) only depends on $\Icn{i}$ and we have defined the gradient operator in the $\mathbb{P}$-plane
\begin{align}
&\nabla_i:=\frac{\partial}{\partial \Icn{i}}\,,&&\forall i=1,2\,.\label{DefNabla}
\end{align}
Hence, for fixed initial conditions, the trajectory of the system in the $\mathbb{P}$-plane can (approximately) be described as a gradient flow. Notice in this regard that $\pot$ (through $(\lambda_i,A_i)$), also implicitly depends on $(\mi_{1,0},\mi_{2,0})$. 

Before moving on to the cases $\sig_{1,2}>1$, in passing we make three more remarks:
\begin{itemize}
\item We find evidence for the fact that $(\lambda_i,A_i)$ depend also on $\sig_i$, but not on $\sig_{j\neq i}$. Concretely, the numerical solutions can be approximated by
\begin{align}
&\lambda_i\sim\tilde{\lambda}_i\,r\,,&&A_i\sim\tilde{A}_i\,r\,\left(1+\frac{1}{\sig_i} W(-S_0\,\sig_i\,e^{-\sig_i})\right)\,,
\end{align} 
where $(\tilde{\lambda}_i,\tilde{A}_i)$ are independent of $\sig_{j\neq i}$.

\item The differential equation (\ref{BetaFunctionLinearDef})
\begin{align}
&\frac{d\Icn{i}}{dt}(t)=\lambda_i\,N\left(1-\frac{\Icn{i}}{N A_i}\right)\,,&&\text{with} &&\Icn{i}(t=0)=\mi_{1,0}\,,
\end{align}
can be solved analytically 
\begin{align}
\Icn{i}(t)=A_i\,N+(\mi_{1,0}-N\,A_i)\,e^{-\frac{\lambda_i}{A_i}\,t}\,.\label{FunctSolFrascati}
\end{align}

\item As we have explained, varying the initial conditions $(\mi_{1,0},\mi_{2,0})$ while keeping $\ms_0$ fixed as in (\ref{CondsInit}) leads to a codimension 1 surface of fixed points. However, by changing at the same time $(\sig_1,\sig_2)$, it is possible to find a family of trajectories that flow to a single point in the $\mathbb{P}$-plane (an example is schematically shown in Figure~\ref{Fig:RGTrajectoriesSingleFlow}). This can be achieved for any choice of $\sig_{1,2}$ and is not limited to $\sig_{1,2}<1$.

\begin{figure}[htbp]
\begin{center}
\includegraphics[width=10.5cm]{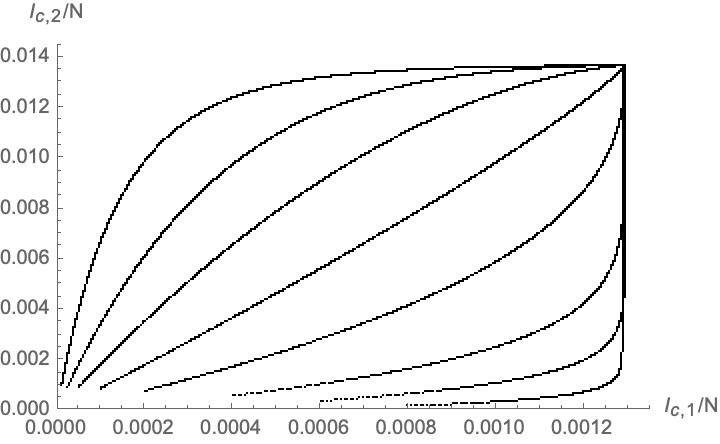}
\end{center}
\caption{Modified trajectories where in addition to the initial condition also $(\sig_1,\sig_2)$ have been modified in such a way to lead to a single point in the $\mathbb{P}$-plane. }
\label{Fig:RGTrajectoriesSingleFlow}
\end{figure}

\end{itemize}

\subsubsection{Flow for $\sig_{1,2}>1$}

\begin{figure}[htbp]
\begin{center}
\includegraphics[width=7.5cm]{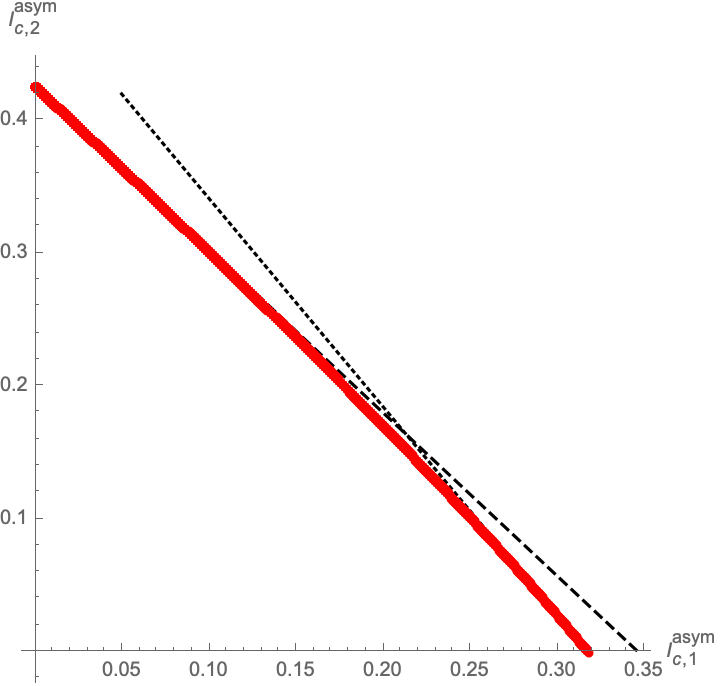}\hspace{1cm} \includegraphics[width=7.5cm]{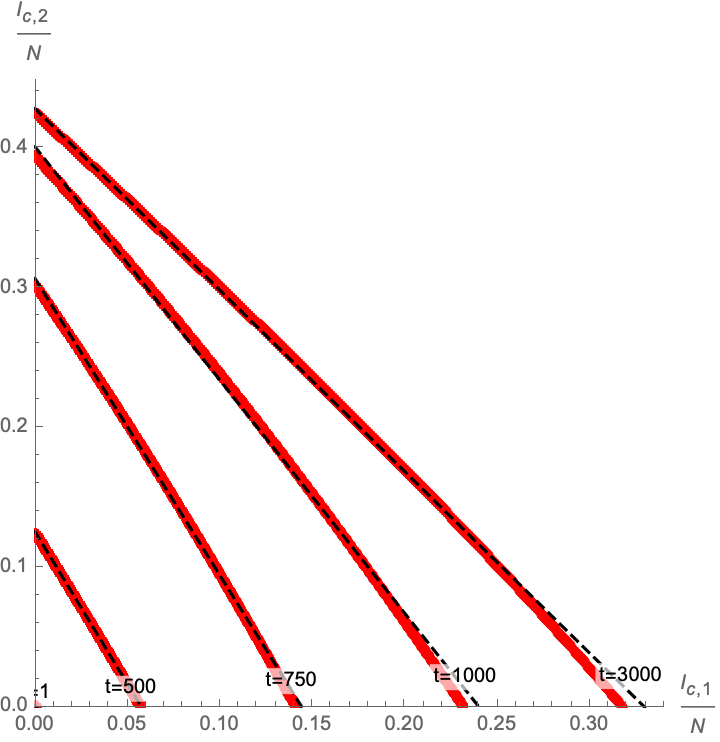}
\end{center}
\caption{Left panel: endpoints of the trajectories of the system in the $\mathbb{P}$-plane for different initial conditions. Each red dot represents $(\Icn{1}^{\text{asym}},\Icn{2}^{\text{asym}})$ at asymptotic time $t\to\infty$ for different choices of $(\mi_{1,0},\mi_{2,0})$ satisfying (\ref{CondsInit}). The black lines correspond to (linear) approximations for initial conditions leading to $\Icn{1,2}^{\text{asym}}\ll 1$ respectively. Right panel: Equal time-slices $(\Icn{1}(t),\Icn{2}(t))$ for finite $t$ (red dots) show deviations from approximate linear relations between $(\Icn{1},\Icn{2})$. Both plots use $\ms_0=0.999$ and $(\sig_1,\sig_2)=(1.2,1.3)$. }
\label{Fig:RGTrajectoriesAsymptoticSup}
\end{figure}

We now move to analysing the case where both $\sigma_{1,2}>1$. The first difference from the previous case is that neither the asymptotic end points $(\Icn{i}^{\text{asym}},\Icn{2}^{\text{asym}})$ (see (\ref{AsymptoticLimitIc}) for the definition) of the trajectories in the $\mathbb{P}$-plane nor equal-time slices for finite $t$ can be approximated by linear relations. As an illustration, Figure~\ref{Fig:RGTrajectoriesAsymptoticSup} shows the corresponding plots for $(\sig_1,\sig_2)=(1.2,1.3)$, highlighting visible deviations from a linear regime. We notice that these deviations become more significant the larger is the difference between $\sig_1$ and $\sig_2$.

\begin{figure}[htbp]
\begin{center}
\includegraphics[width=7.5cm]{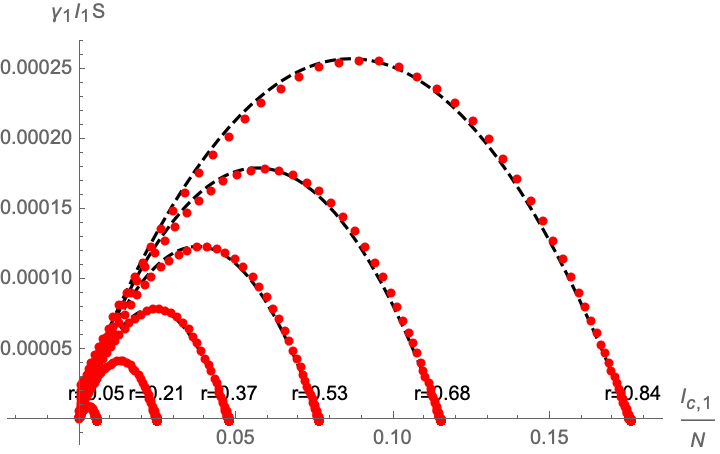}\hspace{1cm} \includegraphics[width=7.5cm]{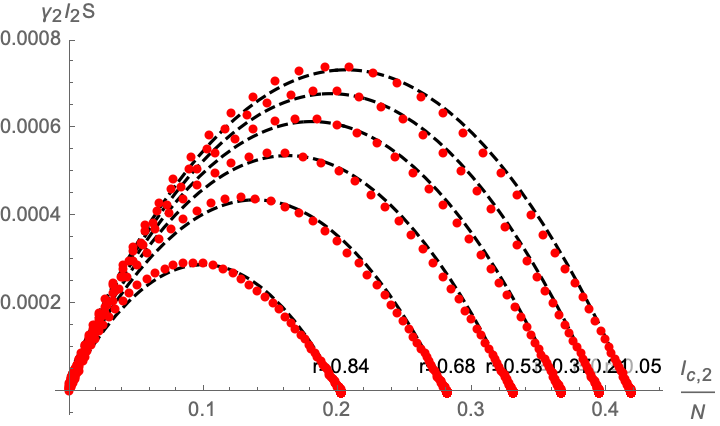}
\end{center}
\caption{Quadratic approximation of $(\Icn{1},\tfrac{d\Icn{1}}{dt})$ (left panel) and $(\Icn{2},\tfrac{d\Icn{2}}{dt})$ (right panel) as a function of the parameter $r$ (defined in eq.~(\ref{DefParamInit})). The red dots represent numerical solutions of $(\Icn{i},\tfrac{d\Icn{i}}{dt})$ for different values of time $t$, grouped according to the initial conditions. All plots use $\ms_0=0.999$ and $(\sig_1,\sig_2)=(1.2,1.3)$.}
\label{Fig:QuadBetaFunctionsSIIRInf}
\end{figure}

We now turn to the differential equations (\ref{DiffEqIcn}) for $\Icn{i}(t)$. The main difference compared to the case $\sig_{1,2}<1$ is that the factors $N\,\rin_i\,\mi_i(t)\,\ms_i(t)$ can no longer be approximated by linear functions in $\Icn{i}$. Instead,  we find evidence that the time-derivatives can be approximated by a quadratic function of the form
\begin{align}
-\beta_i(\Icn{j}(t)):=\frac{1}{N}\,\frac{d\Icn{i}}{dt}(t)\sim\lambda_i\,\Icn{i}(t)\,\left(1-\frac{\Icn{i}(t)}{N\,A_i}\right)+\delta_i\,,&&\forall i=1,2\,,\label{BetaQuadratApprox}
\end{align}
where $(\lambda_i,A_i,\delta_i)$ are constants that implicitly depend on the initial conditions $(\mi_{1,0},\mi_{2,0})$ (we remark that $\delta_i\ll1$ is generically a rather small parameter, since it is related to the initial conditions $\mi_{i,0}$). Indeed, examples of the approximations as functions of the initial conditions parametrised by $r$, as defined in Eq.~(\ref{DefParamInit}), are shown in Figure~\ref{Fig:QuadBetaFunctionsSIIRInf}. By varying $\sig_{1,2}$, we find that the quadratic approximation (\ref{BetaQuadratApprox}) becomes less satisfactory the larger is the difference between $\sig_{1,2}$.

\begin{figure}[htbp]
\begin{center}
\includegraphics[width=10.5cm]{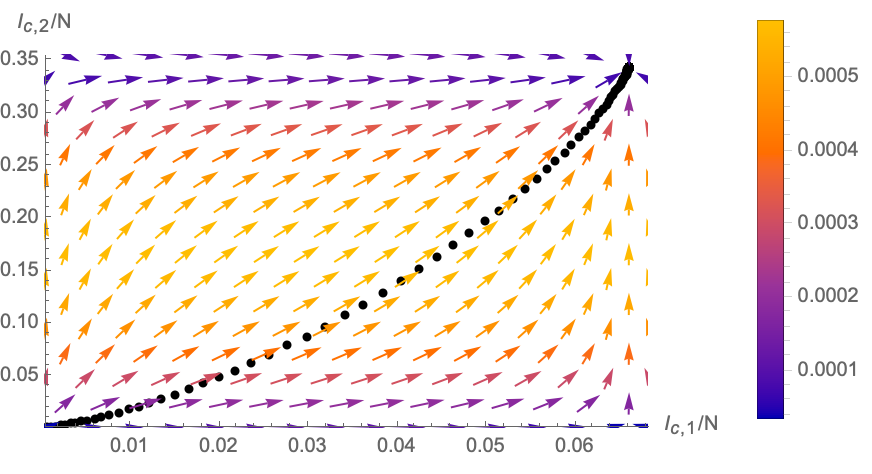}
\end{center}
\caption{Trajectory of the system in the $\mathbb{P}$-plane for a fixed initial condition (black line) following the vector field $\frac{1}{N}(\tfrac{d\Icn{1}}{dt},\tfrac{d\Icn{2}}{dt})$ (the colour of the vectors represents the length). We have used the numerical parameters $r=\tfrac{1}{2}$, $\ms_0=0.999$ and $(\sig_1,\sig_2)=(1.2,1.3)$.}
\label{Fig:RGTrajectoriesQuadDiff}
\end{figure}

As before, we can also represent the trajectory of the system (for fixed initial conditions $(\mi_{1,0},\mi_{2,0})$) in the $\mathbb{P}$-plane as following the vector field $\frac{1}{N}(\tfrac{d\Icn{1}}{dt},\tfrac{d\Icn{2}}{dt})$, as schematically shown in Figure~\ref{Fig:RGTrajectoriesQuadDiff}. Using (\ref{BetaQuadratApprox}), we can therefore write
\begin{align}
&-\beta_i(\Icn{j})=\frac{1}{N}\,\frac{d\Icn{i}}{dt}(t)\sim\nabla_i\,\pot(\Icn{j}(t))\,,&&\text{with}&&\pot(\Icn{j})=\sum_{j=1}^2\left[\Icn{j}^2\,\frac{\lambda_j}{2}\left(1-\frac{2\Icn{j}}{3NA_j}\right)+\delta_j\,\Icn{j}\right]\,,\label{BetaQuad}
\end{align}
where we have used the fact that $\beta_i$ in (\ref{BetaQuadratApprox}) only depends on $\Icn{i}$. The gradient operator in the $\mathbb{P}$-plane $\nabla_i$ is defined in (\ref{DefNabla}). Hence, for fixed initial conditions, the trajectory of the system in the $\mathbb{P}$-plane can be (approximately) described as a gradient flow. Notice in this regard that $\pot$  also implicitly depends on $(\mi_{1,0},\mi_{2,0})$ through $(\lambda_i,A_i)$.

\subsubsection{Flow for $\sig_{1}<1$ and $\sig_2>1$}
We briefly comment on the case where one of the $\sig_i$ (without loss of generality, we can choose it to be $\sig_1$) is smaller than 1 and the other one larger than 1. This corresponds to a situation, where $\mi_1(t)$ is a monotonically decreasing function of time, while $\mi_2(t)$ has a maximum before tending to 0 for large $t$. 

The situation is a combination of the cases discussed in the previous subsections. Notably, the time evolution of $\Icn{1}$ can be approximated by an equation of the form (\ref{BetaLin}) with a function $\pot$ that is quadratic in $\Icn{1}$, while the time evolution of $\Icn{2}$ can be approximated by an equation of the form (\ref{BetaQuad}) with a $\pot$ that is cubic in $\Icn{2}$. We remark that, as before, the approximations are less satisfactory the larger is the difference between $\sig_1$ and $\sig_2$.

\section{Mutation eRG (MeRG) Approach}\label{Sect:RG}

The SIIR model discussed in the previous Section is a simple model that allows to gain basic intuition about the dynamics of two competing variants of a disease, without assuming too much about basic `microscopic' processes that govern its spread. However, due to the simplicity (notably the fact that infection and recovery rates are assumed to be constant in time), this model is not particularly useful to confront (or even predict) real world data. In principle, it is possible to extend the model by allowing for time-dependent $(\rin_i,\rhe_i)$ and/or adding additional compartments, however at the price of complicating the analysis and loosing predicting power.

Hence, in the following we will use the intuition we have gained about the pandemic with two variants to extend the eRG approach. The latter is an economical effective approach that entails a high degree of predictivity in terms of the  time structure of the pandemic.
As discussed in \cite{DellaMorte:2020wlc}, the eRG framework makes use of particular symmetries in the time evolution of an epidemic to give a simplified description of certain key quantities (namely the cumulative number of infected individuals), while more microscopic degrees of freedom have already been taken into account (or `integrated out'). See \cite{Cacciapaglia:2021vvu} for a more detailed description of this philosophy.

\subsection{The Model}

The eRG approach consists in defining $\beta$-functions that govern the time evolution of the system at the global level. To this end, in analogy to particle physics, for each variant we first define an epidemic `coupling strength' as a monotonic, differentiable and bijective, function $\alpha_i=f_i(\Icn{i})$ of the cumulative number of infected individuals. Different choices of $f_i$ correspond to different renormalisation schemes and we expect physical results in general not to depend on the choice. The $\beta$-functions are then defined as
\begin{align}
&-\beta_i=\frac{d\alpha_i}{dt}=\sum_{j=1}^2\frac{df_i}{d\Icn{j}}\,\frac{d\Icn{j}}{dt}(t)\,,&&\forall i=1,2\,.\label{GeneralBetaFunction}
\end{align}
The intuition obtained from the SIIR model suggests to model these $\beta$-functions as polynomials in $\Icn{i}$, which is either linear (for variants with a basic reproduction number $\sigma_i<1$) or quadratic (for variants with a basic reproduction number $\sigma_i>1$). In fact, comparing the equations that emerged in the context of the SIIR model, \emph{i.e.} (\ref{BetaFunctionLinearDef}) with (\ref{BetaQuadratApprox}) (and neglecting\footnote{As we have remarked above, in the SIIR model the (small) parameters $\delta_i$ are related to the (finite) initial conditions $\mi_{i,0}$ of the (active) number of infectious individuals which is necessary to start the time evolution of the system. In the context of the eRG, as we shall discuss in more detail below, we model the point $(\Icn{1},\Icn{2})=(0,0)$ (\emph{i.e.} the absence of the disease) as a fixed point, albeit a repulsive one, which requires $\delta_i=0$.} the small coefficient $\delta_i$), we see that both can in fact be brought into the same framework (\ref{GeneralBetaFunction}) simply by different choices of the functions $f_i$, namely
\begin{align}
f_i(\Icn{i})=\left\{\begin{array}{lcl}c_i\,\ln(\Icn{i}) & \text{for} & \sig_i<1\,, \\ c_i\,\Icn{i} & \text{for} & \sig_i>1\,,\end{array}\right.
\end{align}
for some constant $c_i$. Moreover, (\ref{BetaLin}) and (\ref{BetaQuad}) suggest to further model the beta function in the form of a gradient equation, \emph{i.e.}
\begin{align}
-\beta_i=\frac{d\alpha_i}{dt}=\sum_{j=1}^2\frac{df_i}{d\Icn{j}}\,\frac{d\Icn{j}}{dt}(t)=\nabla_i\,\pot(\Icn{_j})\,.\label{GeneralBetaFunctionGradient}
\end{align}
where the gradient operator is defined in (\ref{DefNabla}) and $\pot$ is a quadratic or cubic function, respectively. In fact, taking these results together with the liberty of scheme-redefinitions and comparing with the general form of the eRG for single variants (see \cite{cacciapaglia2020evidence}) we are naturally led to choose for simplicity $f_i(\Icn{i})=\Icn{i}$ and model $\Phi$ as
\begin{align}
\Phi(\Icn{j})=\sum_{j=1}^2\Icn{j}^2\,\frac{\lambda_j}{2}\left(1-\frac{2\Icn{j}}{3NA_j}\right)\,.\label{PotentialBetaFunction}
\end{align}
In practice, this means that we are modelling the time-evolution of each variant as independent of the other, except for a potential change in the parameters $(\lambda_i,A_i)$. From the intuition obtained from the SIIR model, this is justified as long as the two variants do not have effective reproduction numbers that are largely different from one another (see Section~\ref{Sect:ImpactCumulInfect}). We shall see that this assumption also leads to reasonable results compared to real world data. Indeed, the system (\ref{GeneralBetaFunctionGradient}) for $\pot$ given in (\ref{PotentialBetaFunction}) allows for an analytic solution of $(\Icn{1},\Icn{2})(t)$ which can be written in terms of logistic functions
\begin{align}
&\Icn{i}(t)=\frac{NA_i}{1+e^{-\lambda_i(t-\kappa_i)}}\,,&&\forall i=1,2\,,\label{LogisiticSol}
\end{align}
which well reproduce the data for the first wave of the COVID-19 pandemic \cite{DellaMorte:2020wlc}.
Here $\kappa_{1,2}$ are integration constants that determine which trajectory the system follows in the $\mathbb{P}$-plane. These parameters resemble the initial conditions $(\mi_{1,0},\mi_{2,0})$ from the perspective of the SIIR model.

\subsection{Structure of the $\beta$-Functions in the $\mathbb{P}$-plane}
\begin{figure}[htb]
\begin{center}
\includegraphics[width=10.5cm]{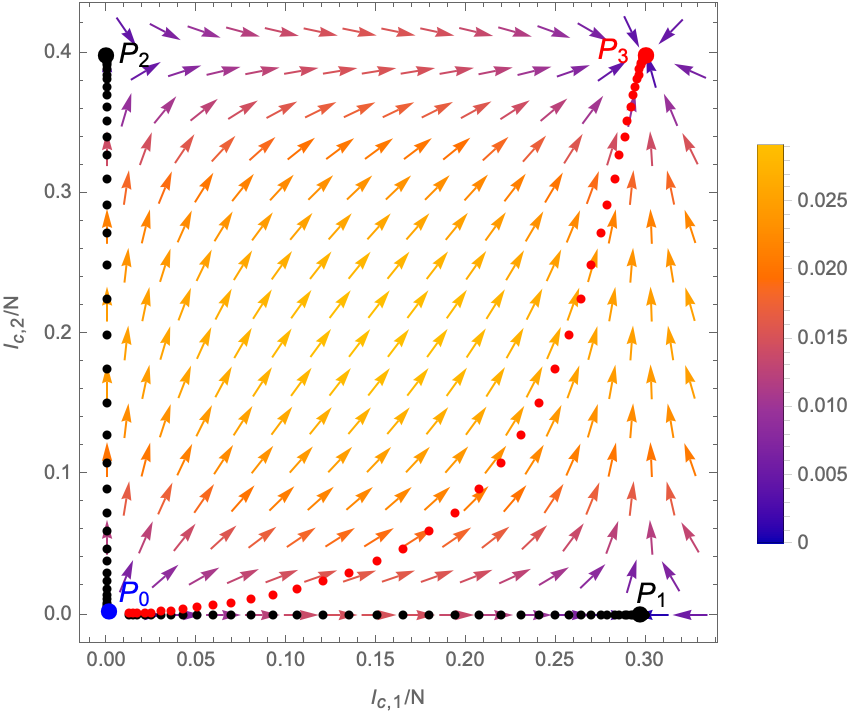}
\end{center}
\caption{Schematic structure of the RG-equations in the $\mathbb{P}$-plane (the values used for the plot are $N\lambda_1=0.2$, $N\lambda_2=0.25$ and $A_1=0.3$, $A_2=0.4$). The $\beta$-functions exhibit four fixed points: $P_{0,1,2,3}$ of which $P_0$ is repulsive in all directions, $P_{1,2}$ have one attractive direction and one repulsive one, while $P_3$ is attractive in all directions. Different trajectories as solutions of (\ref{GeneralBetaFunctionGradient}) connecting the fixed points are indicated by the dotted lines. The colouring of the vectors indicates the norm~$||(\beta_1,\beta_2)||$, which respectively leads to smaller or larger distances between different points of the flow lines. }
\label{Fig:RGFlowFixedPoints}
\end{figure}

The vector field $(- \beta_1(\Icn{i}), - \beta_2(\Icn{i}))$ in the $\mathbb{P}$-plane is schematically plotted in Figure~\ref{Fig:RGFlowFixedPoints}. It has four fixed points, \emph{i.e.} points $(\Icn{1}^{(k)},\Icn{2}^{(k)})$ (for $k=0,1,2,3$) where
\begin{align}
&\beta_i(\Icn{1}^{(k)},\Icn{2}^{(k)})=0\,,&&\forall i=1,2\,,
\end{align}
explicitly given by
\begin{align}
&P_0=(\Icn{1}^{(0)},\Icn{2}^{(0)})=(0,0)\,,&&P_1=(\Icn{1}^{(1)},\Icn{2}^{(1)})=(N A_1,0)\,,\nonumber\\
&P_2=(\Icn{1}^{(2)},\Icn{2}^{(2)})=(0,N A_1)\,,&&P_3=(\Icn{3}^{(k)},\Icn{2}^{(3)})=(N A_1,N A_2)\,.
\end{align}

\noindent
Among them, $P_0$ is repulsive in all direction (\emph{i.e.} in Figure~\ref{Fig:RGFlowFixedPoints} all arrows point away from it) and corresponds to the case where no disease is (and never has been) present. In fact, moving away from this fixed point by infecting even only a small number of individuals of the population (with either of the two variants) causes the system to flow to one of the other three fixed points. Among them, $P_{1,2}$ are repulsive in one direction, but attractive in the other. Since they are characterised by $\Icn{2}=0$ or $\Icn{1}=0$ respectively, they correspond to the endpoints of scenarios in which variant 2 or variant 1 is never present in the population (\emph{i.e.} all infected individuals are infected with only one of the two variants). These fixed points can be reached only by the flow lines represented in black in Figure~\ref{Fig:RGFlowFixedPoints}, which are initiated by a deformation away from $P_0$ along $\Icn{1}$ or $\Icn{2}$ only, respectively.\footnote{We shall discuss in the following subsection~\ref{Sect:CriticalSurfaceMutation} in more detail the case in which along one of the black trajectories in Figure~\ref{Fig:RGFlowFixedPoints} infectious individuals of the other variant appear. This scenario models for example the appearance of a mutation of an already present variant. In this case the system will flow to the fixed point $P_3$ (rather than continue towards $P_1$ or $P_2$.)} Any deformation that switches on both $\Icn{1}$ and $\Icn{2}$ (\emph{i.e.} any scenario in which infected with both variants are present in the population) causes the system to flow to fixed point $P_3$, an example of such a flow is indicated in red in Figure~\ref{Fig:RGFlowFixedPoints}. Which trajectory is realised depends on the initial deformation, which is represented by the parameters $\kappa_{1,2}$ in the solution (\ref{LogisiticSol}) (and which is akin to the choice of different initial conditions in the case of the SIIR model).


\subsection{Critical Surfaces and Mutations}\label{Sect:CriticalSurfaceMutation}
The assumption of a deformation away from the fixed point $P_0$ in Figure~\ref{Fig:RGFlowFixedPoints} (which represents the complete absence of the disease in the population) along a generic direction is not realistic, since it would correspond to the simultaneous appearance of infected individuals of both variants in the population. A more likely scenario would be the appearance of one variant first, while a second deformation at a latter stage introduces the second variant. This dynamics can be understood from an RG perspective as the switching on of a relevant operator.

To make this statement more precise, we first need to introduce the concept of \emph{critical surface} associated with a fixed point of the $\beta$-functions. A critical surface consists in all points in the $\mathbb{P}$-plane from where the RG-flow leads to the fixed point in question. Concretely, for the fixed points $P_{1,2,3}$, the critical surfaces are
\begin{align}
\mathcal{C}_{P_1}=\{(\Icn{1},\Icn{2})\in\mathbb{P}|\Icn{1}>0\text{ and }\Icn{2}=0\}\,,\nonumber\\
\mathcal{C}_{P_2}=\{(\Icn{1},\Icn{2})\in\mathbb{P}|\Icn{1}=0\text{ and }\Icn{2}>0\}\,,\nonumber\\
\mathcal{C}_{P_3}=\{(\Icn{1},\Icn{2})\in\mathbb{P}|\Icn{1}>0\text{ and }\Icn{2}>0\}\,.
\end{align}
A \emph{relevant operator} (from the perspective of the fixed point in question), corresponds to a direction that drives the theory away from the critical surface, such that it flows to a new critical point. In the case at hand, $\mathcal{C}_{P_{1,2}}$ have one critical direction orthogonal to it, which, from an epidemiological perspective, precisely corresponds to the appearance of the second variant. A small deformation at any point of $\mathcal{C}_{P_{1,2}}$ (for example due to a relevant mutation of the virus) causes the system to deviate from the critical surface and ultimately flow towards $P_3$.

\begin{figure}[htb]
\begin{center}
\includegraphics[width=7.5cm]{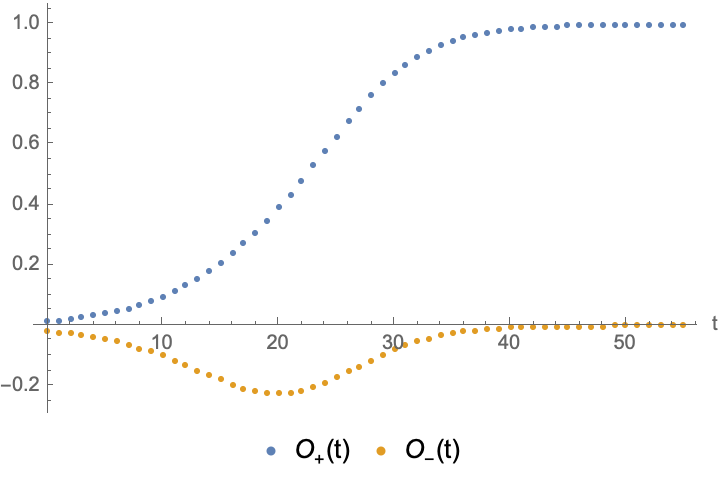}\hspace{1cm} \includegraphics[width=7.5cm]{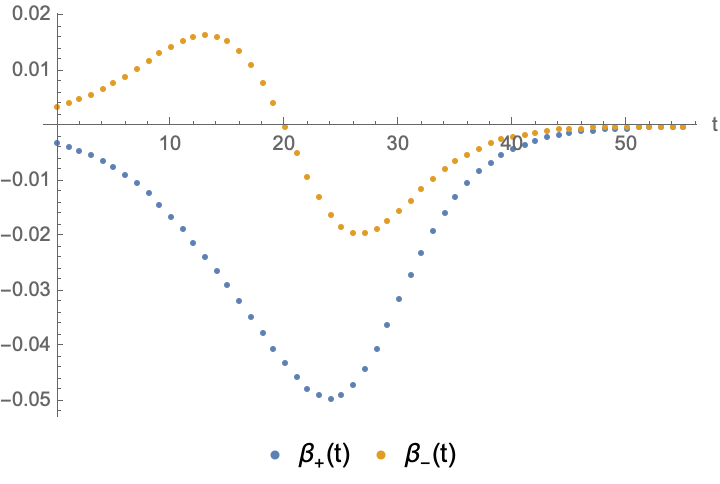}
\end{center}
\caption{Parametric Plot of the operators $\mathcal{O}_\pm$ (left panel) and their $\beta$-functions (right panel) as functions of time. The plots use $N\lambda_1=0.2$, $N\lambda_2=0.25$, $A_1=0.3$, $A_2=0.4$ and $\kappa_1=16$, $\kappa_2=25$.}
\label{Fig:IrrelevantOp}
\end{figure}

In a scenario with only two variants, the fixed point $P_3$ has no relevant deformations and is attractive along all directions. Instead, small fluctuations along trajectories leading towards $P_3$ (such as the red path shown in Figure~\ref{Fig:RGFlowFixedPoints}) can be interpreted as \emph{irrelevant} operators being switched on. To make this statement more concrete, instead of $(\Icn{1},\Icn{2})$ we can consider the following functions\footnote{SH would like to thank Michele Della Morte for useful exchanges on the form of these functions as $O(2)$ transformations of $(\Icn{1},\Icn{2})$.}
\begin{align}
&\mathcal{O}_+=\frac{A_1A_2}{A_1^2+A_2^2}\left(\frac{\Icn{1}}{NA_2}+\frac{\Icn{2}}{NA_1}\right)\,,  &&\text{and} &&\mathcal{O}_-=\frac{A_1A_2}{A_1^2+A_2^2}\left(-\frac{\Icn{1}}{N A_1}+\frac{\Icn{2}}{NA_2}\right)\,.\label{OpmOperators}
\end{align}
Notice that, at the fixed point $P_3$, we have 
\begin{align}
\mathcal{O}_-(\Icn{1}=NA_1,\Icn{2}=NA_2)=0\,. 
\end{align}
Using the explicit solutions of $(\Icn{1},\Icn{2})(t)$ in (\ref{LogisiticSol}) we have schematically plotted $\mathcal{O}_\pm$ and its $\beta$-functions as functions of time in Figure~\ref{Fig:IrrelevantOp}. The latter can be written in the form 
\begin{align}
&-\beta_\pm=\frac{d\mathcal{O}_\pm}{dt}=\frac{1}{N\,(A_1^2+A_2^2)}\,\nabla_\pm\,\Phi(\mathcal{O}_\pm)\,,\text{with} &&\nabla_\pm=\frac{\partial}{\partial\mathcal{O}_\pm}\,.\label{GradientVectorFieldOpm}
\end{align}
We can then interpret the time evolution of the system as a flow of $\op{+}$ from the repulsive fixed point $\op{+}=0$ to the attractive fixed point $\op{+}=1$, while $\op{-}$ can be interpreted as an irrelevant operator that is switched on along the way, which does not drive the system away from the fixed point $(\op{+},\op{-})=(1,0)$.

\subsection{Multi Wave Dynamics and CeRG}
\begin{figure}[htb]
\begin{center}
\includegraphics[width=7.5cm]{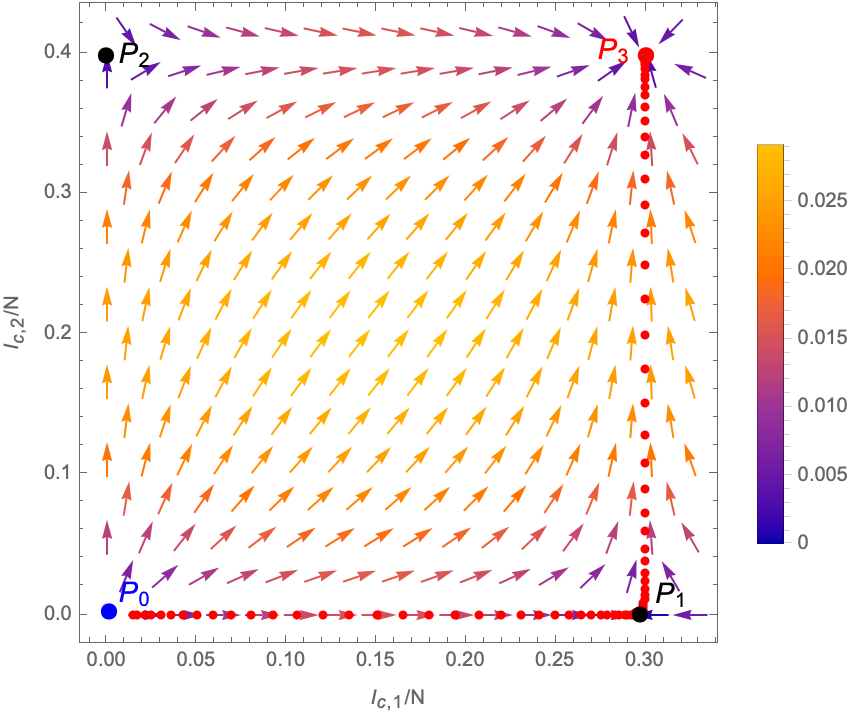}\hspace{1cm} \includegraphics[width=7.5cm]{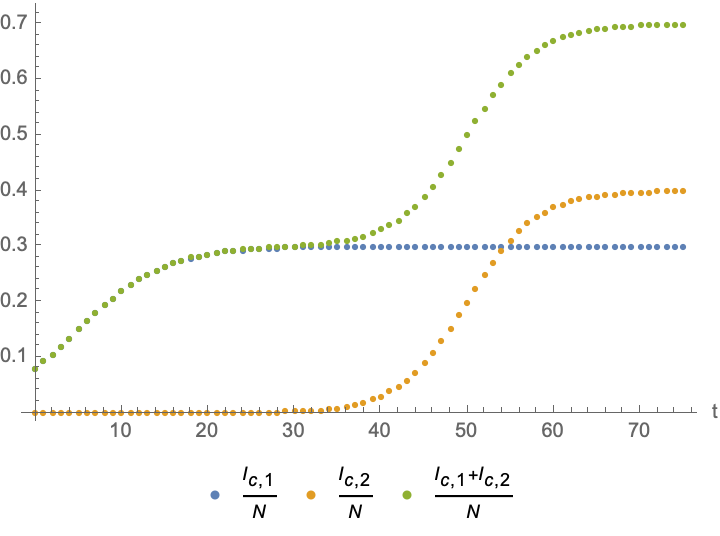}
\end{center}
\caption{Left panel: Schematic example of a crossover flow. During the first part, the system follows closely the critical surface $\mathcal{C}_{P_1}$ (\emph{i.e.} parallel to the $\Icn{1}$-axis) and comes close to the fixed point $P_1$ without reaching it. After staying for some time in the vicinity of $P_1$ the system enters into the second part of the flow to the fixed point $P_3$. Right panel: Cumulative numbers of infected as functions of time. The total number of infected shows a two-wave structure with a linear growth phase at around $t=30$ corresponding to the time that the RG flow is in proximity to $P_1$. The numerical values used for the plots are $N\lambda_1=0.2$, $N\lambda_2=0.25$ and $A_1=0.3$, $A_2=0.4$. }
\label{Fig:RGCrossoverFlow}
\end{figure}

We now consider in more detail a particular case of a flow along the critical surface $\mathcal{C}_{P_1}$ with a relevant operator being switched on along the way (\emph{i.e.} the second variant appearing at some moment $t_0>0$) through a small fluctuation. If $\sig_2$ is not too large compared to $\sig_1$, for some time after the relevant deformation is turned on, the theory stays close to the (initial) critical surface $\mathcal{C}_{P_1}$ and flows towards the fixed point $P_1$. At some later time, the relevant deformation along direction $\Icn{2}$ becomes too large and the flow runs significantly away from the critical surface $\mathcal{C}_{P_1}$. If the deformation appears well after $\mi_1(t)$ has reached a maximum, the flow can be described as a \emph{crossover} flow (see Figure~\ref{Fig:RGCrossoverFlow} for a schematic example): the RG-flow can be decomposed into a flow along the original critical surface $\mathcal{C}_{P_1}$ followed by a flow perpendicular to it. The latter drives the system from the proximity of the fixed point $P_1$ to the new one $P_3$. From the perspective of the new fixed point, the second phase of the flow looks like an RG flow from a UV fixed point $P_1$ to an IR one $P_3$.

Since during the first part of the flow, the number of active infected of the second variant $\mi_2(t)$ is fairly small, this flow is very well approximated by a usual eRG dynamics (see eq.~(\ref{GeneralBetaFunctionGradient})), which has been shown \cite{DellaMorte:2020wlc} to describe real world data very well. Once the system reaches the vicinity of the fixed point $P_1$, it will then enter into a quasi-linear growth phase, in which the number of active infected with respect to both variants is small and therefore the total number of infected $(\Icn{1}+\Icn{2})(t)$ only grows linearly (see Figure~\ref{Fig:RGCrossoverFlow}). However, after a certain time, the number of infected $\Icn{2}$ will grow exponentially (while the number of infected with respect to the original variant remains small) and the system enters into the crossover phase. Now, the $\beta$-function for $\Icn{2}$ is, once again, essentially modelled by a standard eRG equation (see eq.~(\ref{GeneralBetaFunctionGradient})) describing the flow to $P_3$.

In this picture, the two-wave structure is explained as the (more or less successive) appearance of two different variants of the disease. In particular, the linear-growth phase (that has for example been observed in real world data in the inter-wave period of the COVID-19 pandemic \cite{cacciapaglia2020multiwave}) is explained by the fact that the system comes close to a fixed point, which, however, it cannot reach. It nevertheless spends significant time in its proximity. A similar reasoning underlies the \emph{Complex epidemic Renormalisation Group} (CeRG) approach \cite{cacciapaglia2020evidence}.  The CeRG $\beta$-function of the following type was proposed for the combined number of cumulative infected $\Icn{\text{tot}}=\Icn{1}+\Icn{2}$
\begin{align}
-\beta_{\text{CeRG}}=\frac{d\Icn{\text{tot}}}{dt}(t)=\lambda\,\Icn{\text{tot}}(t)\,\left[\left(1-\zeta\,\frac{\Icn{\text{tot}}(t)}{A}\right)^2-\delta\right]^{p_1}\,\left(1-\frac{\Icn{\text{tot}}(t)}{A}\right)^{p_2}\,,\label{CeRG}
\end{align}
with $A$ the asymptotic number of infected, $\lambda$ the infection rate, $\zeta>1$ and $\delta<0$. Indeed, besides the fixed points $\Icn{\text{tot}}=0$ and $\Icn{\text{tot}}=A$, the beta-function (\ref{CeRG}) also has the complex fixed points $\Icn{\text{tot}}=\frac{A}{\zeta}\left(1\pm i\sqrt{|\delta|}\right)$, which cannot be reached by the flow, but are responsible for the linear-growth phase.

\begin{figure}[htbp]
\begin{center}
\includegraphics[width=7.5cm]{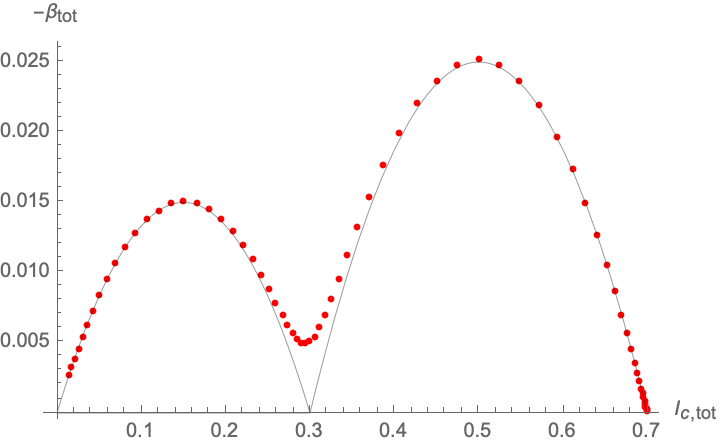}\hspace{1cm}\includegraphics[width=7.5cm]{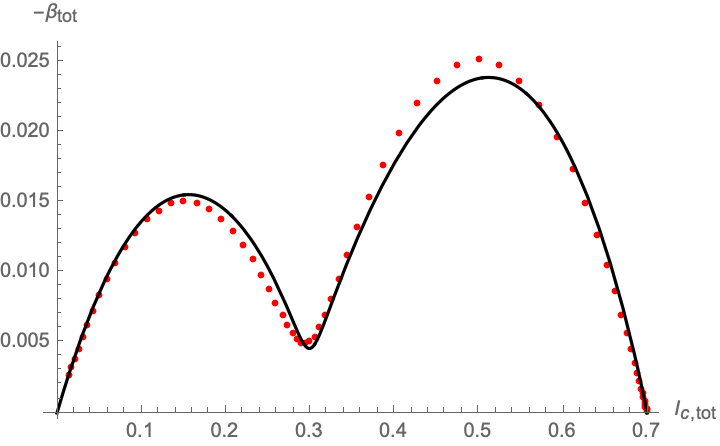}
\end{center}
\caption{Schematic plot of the beta-function of the total number of infected computed from the solutions (\ref{LogisiticSol}) (red dots). Left panel: comparison to (\ref{CombinedTotalBeta}), right panel: fitting with a $\beta$-function of the type~(\ref{CeRG}), with $\kappa=0.2075$, $\delta=-0.0021$, $p_1=0.336$, $p_2=0.959$, $A=0.6996$ and $\frac{A}{\zeta}=0.2995$. }
\label{Fig:RGTotSchem}
\end{figure}

To compare (\ref{GeneralBetaFunctionGradient}) and (\ref{PotentialBetaFunction}) with the $\beta$-function in (\ref{CeRG}), we assume that $\sig_1$ and $\sig_2$ are not significantly different from one another and that the mutation occurs significantly after the maximum number of infected of the first variant (such that the condition of a crossover flow is satisfied). Furthermore, we can write the following combined $\beta$-function for the total cumulative number of infected
\begin{align}
-\beta_{\text{tot}}=&\frac{1}{N}\left[\frac{d\Icn{1}}{dt}(t)+\frac{d\Icn{2}}{dt}(t)\right]\sim\theta\left(\left(NA_1-\Icn{1}\right)\right)\lambda_1\,\Icn{1}\left(1-\frac{\Icn{1}}{A_1}\right)\nonumber\\
&+\theta\left(\left(\Icn{2}-\xi\right)\left(N(A_1+A_2)-\Icn{2}\right)\right)\lambda_2\,\left(\Icn{2}-N\,A_1\right)\left(1-\frac{\Icn{2}-NA_1}{NA_2}\right)\,,\label{CombinedTotalBeta}
\end{align}
where $\theta$ is the Heaviside step-function. Furthermore, we can use the solutions (\ref{LogisiticSol}) of $(\Icn{1},\Icn{2})(t)$ as logistic functions to schematically plot (\ref{CombinedTotalBeta}): the left panel of Figure~\ref{Fig:RGTotSchem} shows a parametric plot of $(\Icn{1}(t)+\Icn{2}(t),\tfrac{d\Icn{1}+\Icn{2}}{dt}(t))$ for different values of $t$. The latter are very well approximated by (\ref{CombinedTotalBeta}) (shown by the thin black line), except for a small region around $\Icn{t}\sim A_1$, in which the beta-function does not in fact reach zero, but interpolates between the two terms in (\ref{CombinedTotalBeta}). This region corresponds to a non-trivial interaction between the variants and governs the transition from the first part of the flow (close to the original critical surface) to the crossover flow. It precisely corresponds to the linear growth region in the context of the CeRG: Indeed, a similar shape of the beta-function can also be achieved through a function of the form (\ref{CeRG}), as is shown in the right panel of (\ref{Fig:RGTotSchem}). The region around $\Icn{\text{tot}}\sim A_1$ corresponds to the RG-flow not quite reaching a zero, thus leading to the quasi-linear growth phase.

\section{Explicit Examples from the COVID-19 pandemic}\label{Sect:RealWorld}

In the following we present selected examples of number of infected individuals for different countries during the COVID-19 pandemic. These serve merely as illustration for some of the points we have outlined in the previous sections.


\subsection{Second wave in California}

\begin{figure}[htbp]
\begin{center}
\includegraphics[width=12.5cm]{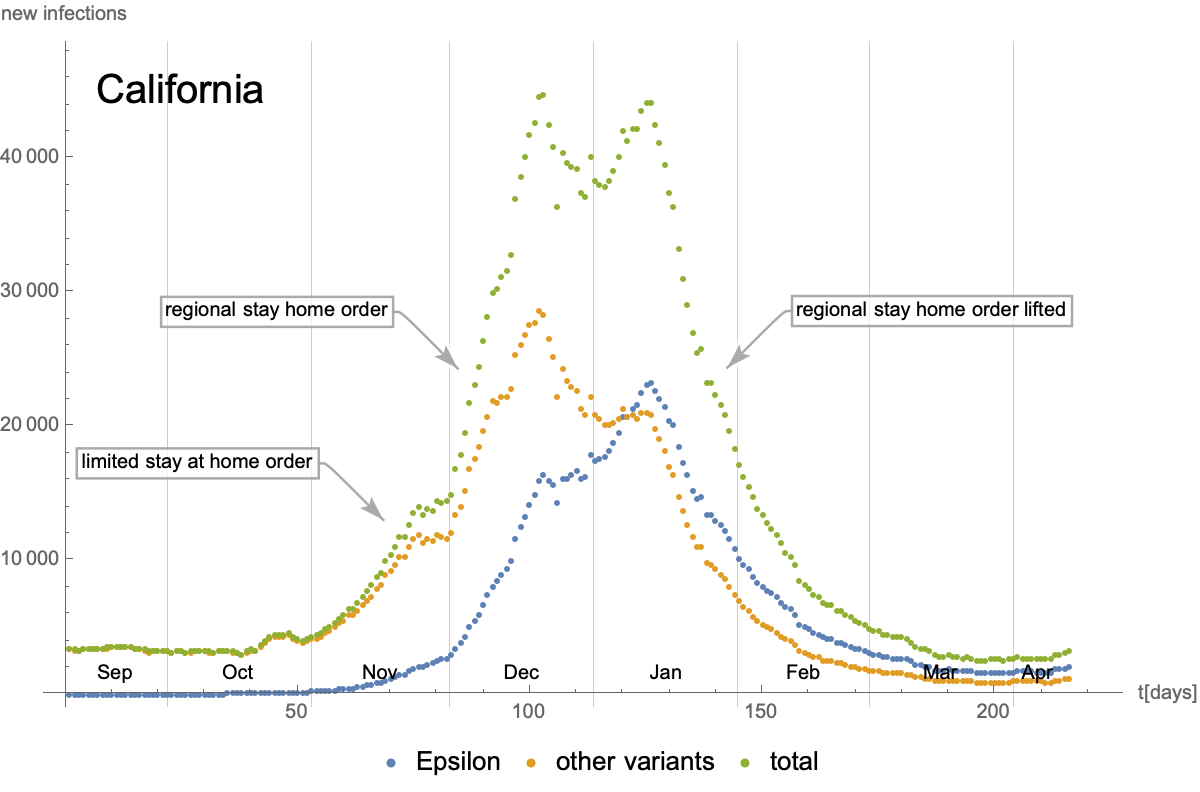}
\end{center}
\caption{Number of daily new infections in California as a function of time since 08/Sep/2020. The epidemiological data have been extracted from \cite{OurWorldInData,NYT}, while the number of cases for the Epsilon variant are based on genome sequencing data extracted from \cite{Gisaid}. Finally, the information about lockdown and social distancing measures is taken from \cite{WikiCal}.}
\label{Fig:California}
\end{figure}

As a first example, we consider the evolution of the SARS-CoV-2 Epsilon\footnote{This variant comprises the variants lineage B.1.429 and B.1.427 under the Phylogenetic Assignment of Named Global Outbreak Lineages (pangolin) tool and was first detected in California in July 2020.} variant relative to the other variants in California. The daily number of new cases is shown in Figure~\ref{Fig:California} and we have also indicated changes in the social distancing and lockdown measures imposed at the same time. To obtain the number of new infections with the Epsilon variant relative to the others, we have used data from GISAID \cite{Gisaid} that provide the sequenced genomes of samples taken in California from Sep/2020 until Apr/2021 (we refer to \cite{Companion} for more details on our methodology): while only a fraction of all positive test samples per day is genetically analysed, we assume that the distribution of the Epsilon variant in this subset is representative of the distribution of the variant among all infected individuals in California. We therefore have first calculated the percentage of each variant among the sequences analysed at each specific date. By multiplying this percentage with the total number of new cases (obtained from \cite{OurWorldInData,NYT}) within all of California, we are able to extract an approximation of the number of new cases per day for each variant. The statistical uncertainty inherent in this procedure has been estimated and taken into account when fitting the cumulative number of cases (see Figure~\ref{Fig:CaliforniaIc} below). Due to the large number of tests and genome sequencings performed in California and the rather short duration of our study, the resulting uncertainty is moderate. Indeed, the maximal number of new infections for both curves in Figure~\ref{Fig:California} occur during a period of 2-3 months (Nov/2020 until Jan/2021), during which furthermore the regional lockdown measures have stayed unchanged. Therefore, in order to model the second wave of the COVID-19 pandemic in California, our implicit assumption that the parameters of the eRG model (\emph{i.e.} $\lambda_{1,2}$ and $A_{1,2}$) are time-independent seems a reasonable starting point. Indeed, from the data in Figure~\ref{Fig:California} we can calculate the cumulative number of infected individuals, which is shown in Figure~\ref{Fig:CaliforniaIc} along with an approximation in terms of logistic functions (\ref{LogisiticSol}) with constant coefficients. Denoting the cumulative number of individuals infected with the Epsilon variant and the other ones by $\Icn{2}$ and $\Icn{1}$ respectively, the parameters extracted from the fit are given by

\begin{figure}[htbp]
\begin{center}
\includegraphics[width=12.5cm]{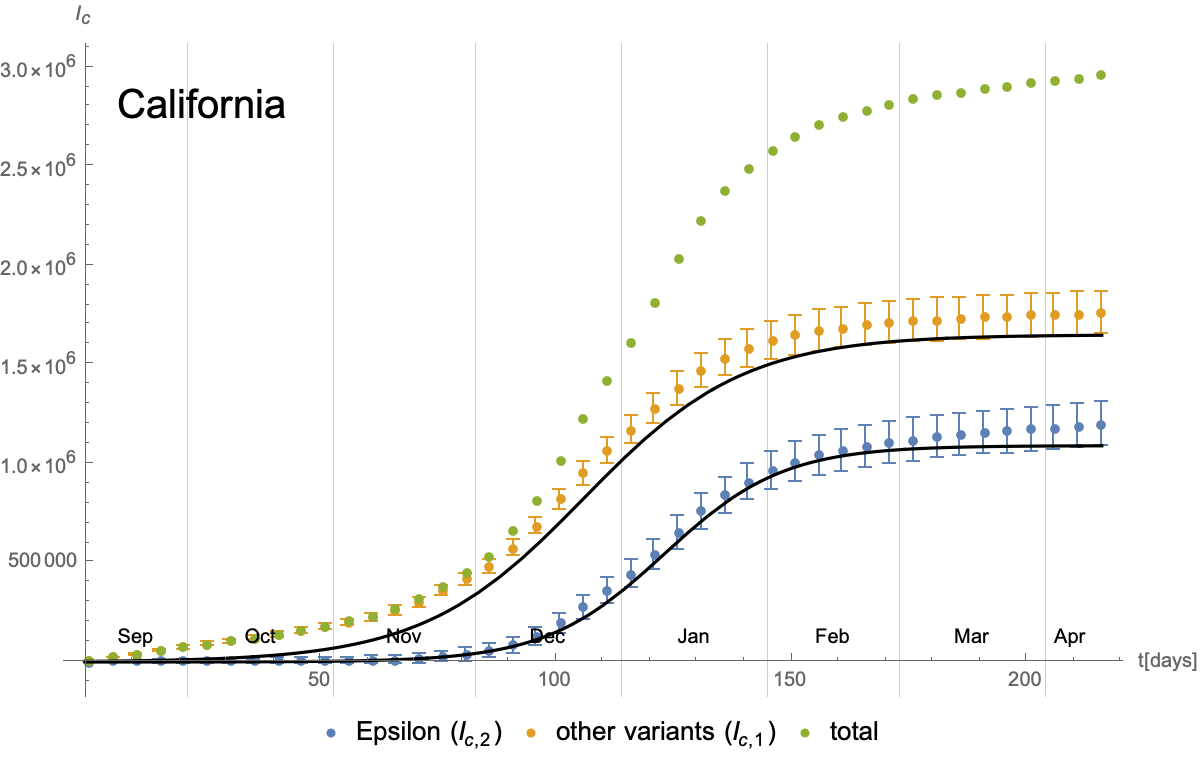}
\end{center}
\caption{Cumulative number of infected individuals with respect to the different variants in California since 08/Sep/2020 and their approximation in terms of logistic functions. The latter have in fact been optimised to fit the daily new infections taking into account their statistical errors.}
\label{Fig:CaliforniaIc}
\end{figure}

\begin{figure}[htbp]
\begin{center}
\includegraphics[width=7.5cm]{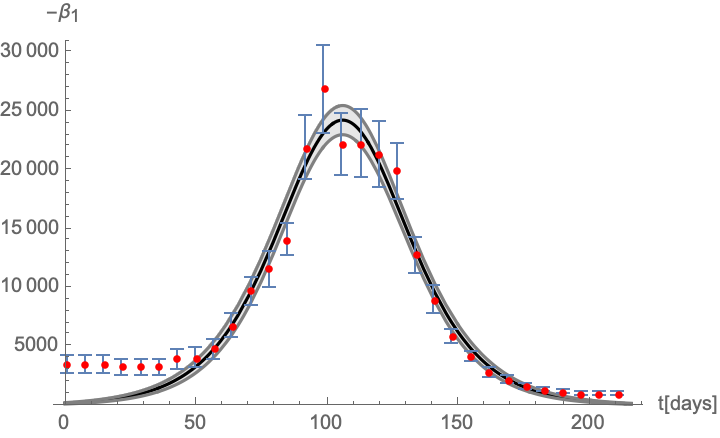}\hspace{1cm}\includegraphics[width=7.5cm]{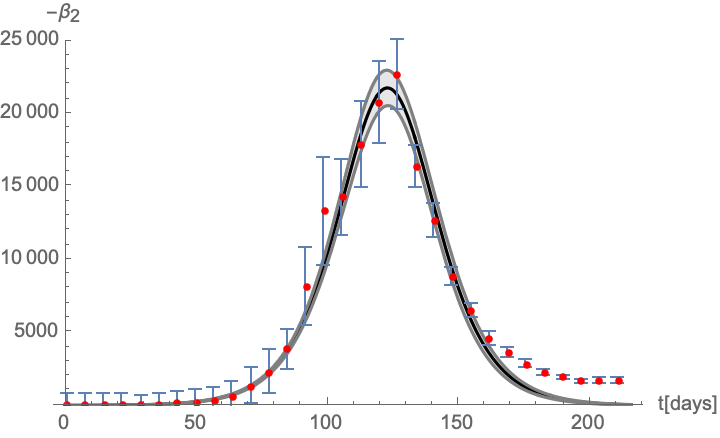}
\end{center}
\caption{Comparison of the time-dependence of the $\beta$-functions (\ref{GeneralBetaFunctionGradient}) for $(\Icn{1},\Icn{2})$ with the numbers of daily new infections. }
\label{Fig:CaliforniaQualTime}
\end{figure}

\begin{figure}[htbp]
\begin{center}
\includegraphics[width=7.5cm]{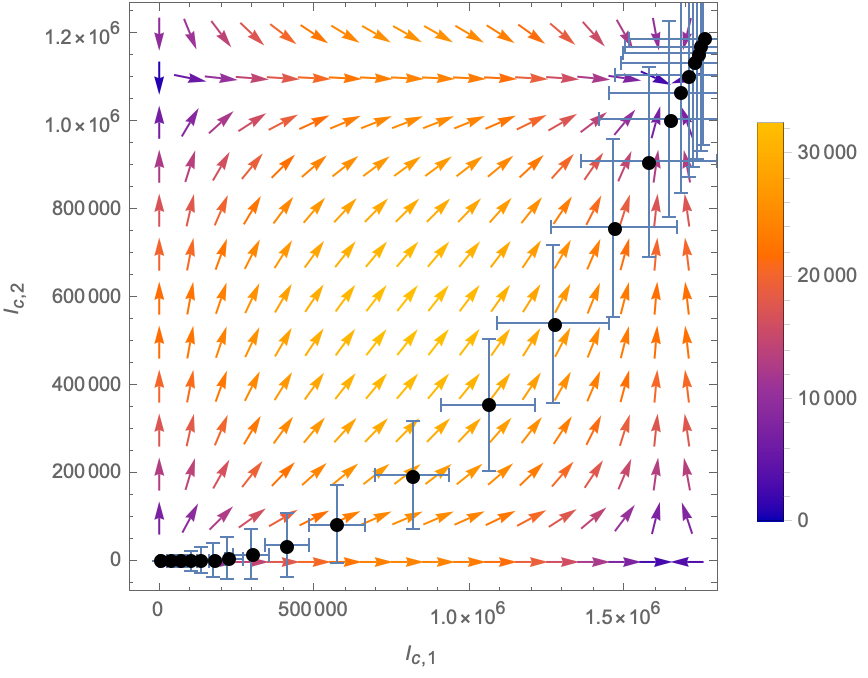}\hspace{1cm}\includegraphics[width=7.5cm]{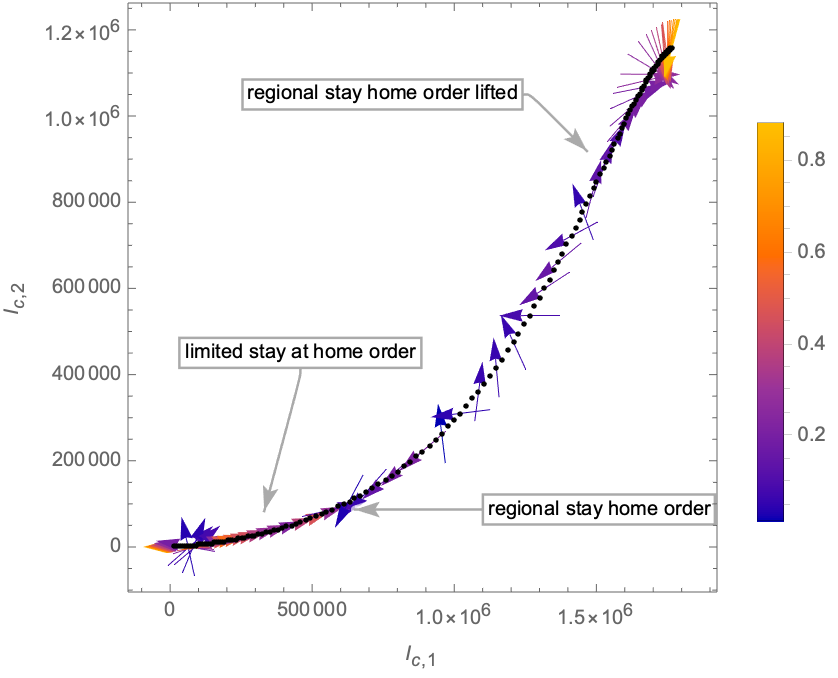}
\end{center}
\caption{Left panel: flow of the system in the $(\Icn{1},\Icn{2})$-plane. The vectors indicate the $\beta$-function as given by  (\ref{GeneralBetaFunctionGradient}) and (\ref{PotentialBetaFunction}) (with $f_i(\Icn{i})=\Icn{i}$ for $i=1,2$). Right panel: relative difference of the modelled $\beta$-function (\ref{GeneralBetaFunctionGradient}) with the actual number of daily infected along the flow of the system in the $(\Icn{1},\Icn{2})$-plane.}
\label{Fig:CaliforniaFlow}
\end{figure}

{\allowdisplaybreaks
\begin{align}
&NA_1=(1.644 \pm 0.021)\cdot 10^6\,,&&\kappa_1=105.92\pm0.40\,,&&\lambda_1=0.0590\pm0.0009\,,\nonumber\\
&NA_2=(1.090\pm 0.014)\cdot 10^6\,,&&\kappa_2=123.00\pm 0.32\,,&&\lambda_2=0.0799\pm 0.0007\,.\label{CalFits}
\end{align}}
These numbers are in fact determined by fitting the derivative of $\Icn{1,2}$ (\emph{i.e.} the daily new cases), as shown in Figure~\ref{Fig:CaliforniaQualTime}. With the help of these approximations, we can compute the function $\pot$ in (\ref{PotentialBetaFunction}) which in turn allows us to compute the $\beta$-functions (\ref{GeneralBetaFunctionGradient}). The corresponding flow of the system in the $(\Icn{1},\Icn{2})$-plane is shown in the left panel of Figure~\ref{Fig:CaliforniaFlow}, where we have chosen the scheme $f_i(\Icn{i})=\Icn{i}$. The right panel of that figure gives the relative deviation of the $\beta$-functions from the time derivative of $(\Icn{1},\Icn{2})$ as given by the number of daily new cases. The plot also indicates changes imposed by the government on the social distancing measures among the population, leading to a stronger deviation of the modelled $\beta$-functions from the actual data.

\begin{figure}[htbp]
\begin{center}
\includegraphics[width=7.5cm]{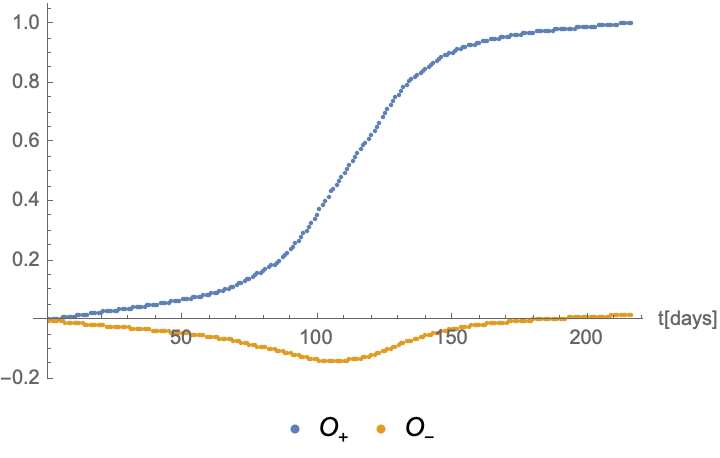}\hspace{1cm}\includegraphics[width=7.5cm]{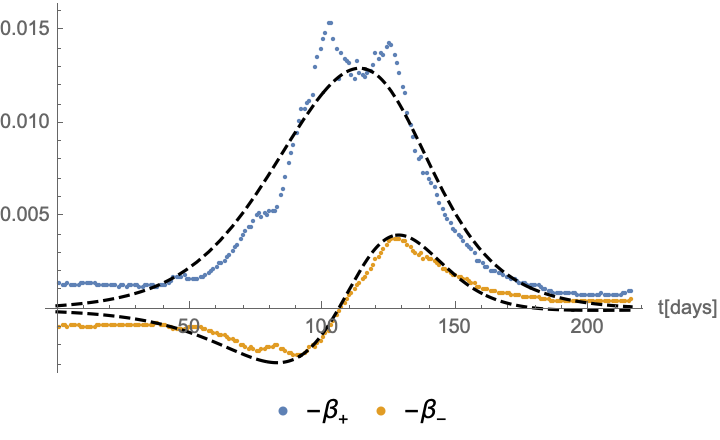}
\end{center}
\caption{Left panel: The functions $\mathcal{O}_\pm$ defined in eq.~(\ref{OpmOperators}) computed from the actual numbers of infected individuals as a function of time. Right panel: The corresponding $\beta$-functions together with their approximations (dashed black lines) implied by (\ref{CalFits}).}
\label{Fig:CaliforniaOpOom}
\end{figure}

\begin{figure}[htbp]
\begin{center}
\includegraphics[width=10.5cm]{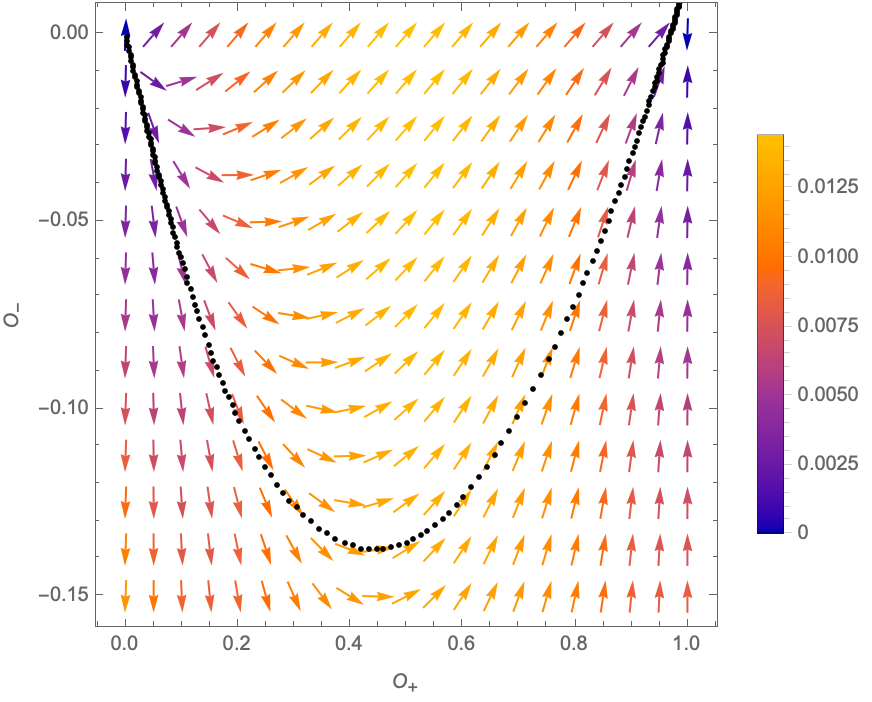}
\end{center}
\caption{Flow of the system in the $(\mathcal{O}_+,\mathcal{O}_-)$-plane based on the data for California. }
\label{Fig:CalFlowDiag}
\end{figure}

\noindent
From the left panel of Figure~\ref{Fig:CaliforniaFlow} we can see that the flow of the system for the most part is not close to either of the two axes, and thus has not the markings of a crossover flow. This is also clear from the fact that the maxima of daily new infections of the variant B.1.427 and the remaining ones are not very well separated in time (see Figure~\ref{Fig:California}). Nevertheless, we expect that the formulation of the flow in terms of the relevant operator $\mathcal{O}_+$ and the irrelevant operator $\mathcal{O}_-$ defined in Eq.~(\ref{OpmOperators}) may still give a viable description. The time dependence of $\mathcal{O}_\pm$ is shown in the left panel of Figure~\ref{Fig:CaliforniaOpOom}, while the right panel of the same figure shows an approximation of the corresponding $\beta$-functions. Finally, the flow of the system in the $(\mathcal{O}_+,\mathcal{O}_-)$ plane (along with the gradient vector field (\ref{GradientVectorFieldOpm})) is shown in Figure~\ref{Fig:CalFlowDiag}.

We remark that the main reason for deviation of the flow from a gradient flow (as is showcased in the right panel of Figure~\ref{Fig:CaliforniaFlow}) can be attributed to the fact that the active number of infected individuals is not zero at the beginning and the end of the flow (see Figure~\ref{Fig:CaliforniaQualTime}). This is, in fact, because we are not describing the entire pandemic in California, but only the period of September 2020 to April 2021, when a second wave hit the state. Therefore, we are not really describing the flow from one fixed point to another, but rather the flow between two linear-growth phases of the system. As such, the description in terms of the $\beta$-function (\ref{GeneralBetaFunctionGradient}) is only approximative, which is particularly visible in the beginning and the end of the flow.

\subsection{Second and third waves in the UK}
\begin{figure}[htbp]
\begin{center}
\includegraphics[width=12.5cm]{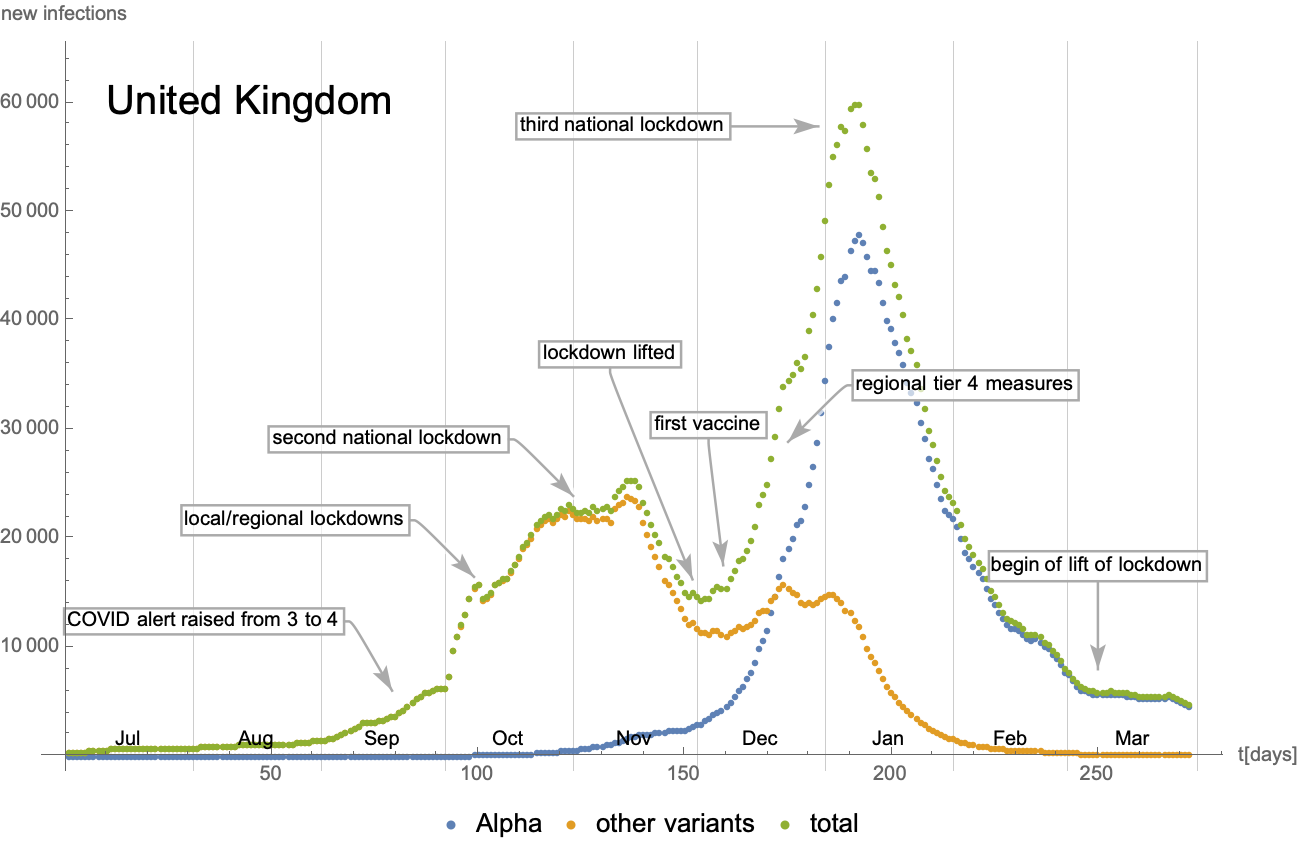}
\end{center}
\caption{Number of daily new infections in the United Kingdom as a function of time since 01/July/2020. The epidemiological data have been extracted from \cite{OurWorldInData}, while the number of cases for the Alpha variant are based on genome sequencing data extracted from \cite{Gisaid}. Finally, the information about lockdown and social distancing measures is taken from \cite{WikiUK}.}
\label{Fig:UK}
\end{figure}

\begin{figure}[htbp]
\begin{center}
\includegraphics[width=12.5cm]{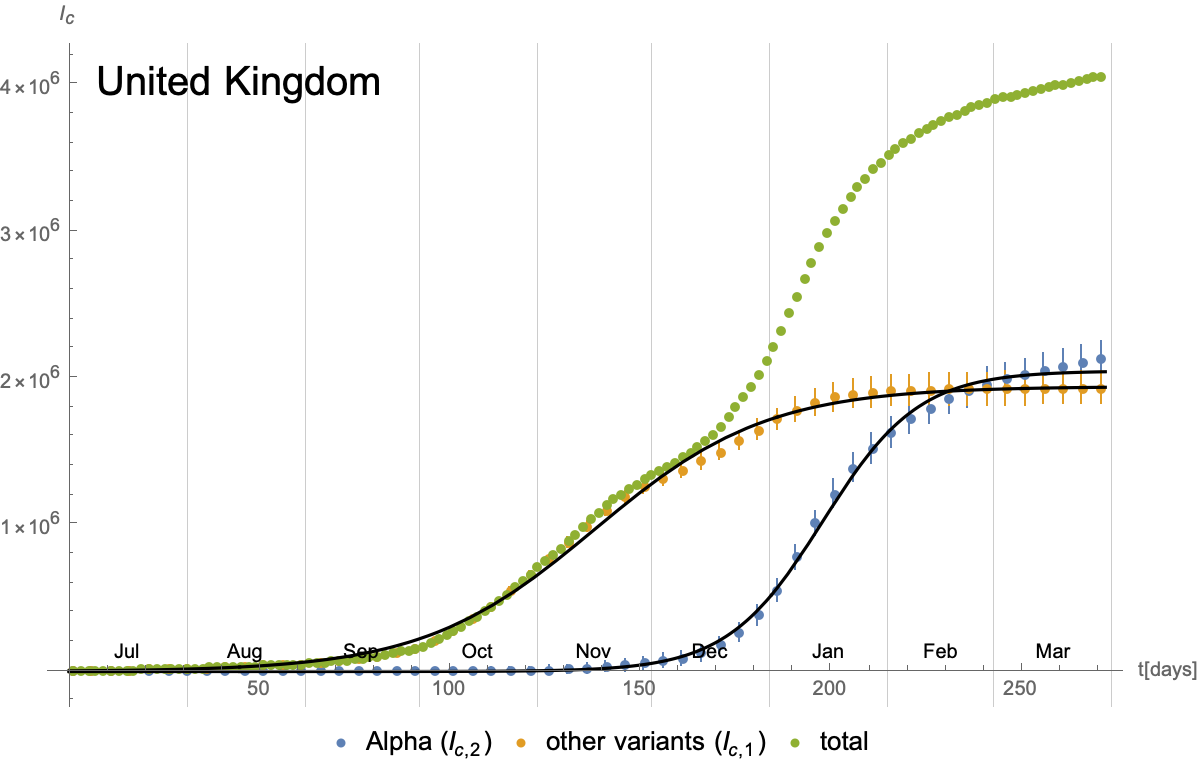}
\end{center}
\caption{Cumulative number of infected individuals with respect to the different variants in the United Kingdom since 01/July/2020 and their approximation in terms of logistic functions. The latter have in fact been optimised to fit the daily new infections.}
\label{Fig:UKIc}
\end{figure}

As a next example, we consider the time evolution of the SARS-CoV-2 Alpha\footnote{This variant is also called lineage B.1.1.7 under the pangolin tool and was first found in Nov. 2020 (in a sample dating from September 2020) in the UK.} variant in the United Kingdom. The daily number of new cases starting from 01/July/2020 is shown in Figure~\ref{Fig:UK}, where we have also indicated changes in the social distancing and lockdown measures. The number of new infections with the Alpha variant have been extracted using GISAID \cite{Gisaid} combined with epidemiological data from \cite{OurWorldInData}. Due to the high number of PCR tests and genome analysis performed in the UK, the inherent statistical uncertainty is rather small (see \emph{e.g.} Figure~\ref{Fig:UKIc}). In contrast to the evolution (of a single wave) in California in the previous subsection, the time evolution studied in this example spans two distinct waves lasting roughly 5 months (Oct/2020 until Feb/2021). As Figure~\ref{Fig:UK} indicates, during this period\footnote{We also remark that the time period includes the Christmas holidays, which traditionally leads to increased social activity and travel among the population}, the lockdown measures have not remained constant, which potentially leads to additional effects, as we shall remark later on. From Figure~\ref{Fig:UK} we can compute the cumulative number of infected individuals, which is shown in Figure~\ref{Fig:UKIc}. The latter also shows approximations in terms of logistic functions: denoting the cumulative number of individuals infected with the Alpha variant and the other ones by $\Icn{2}$ and $\Icn{1}$ respectively, the parameters extracted from the fit are given by

\begin{figure}[htbp]
\begin{center}
\includegraphics[width=7.5cm]{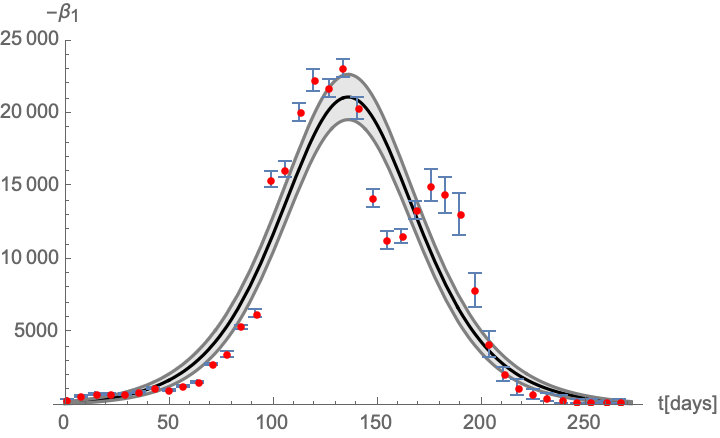}\hspace{1cm}\includegraphics[width=7.5cm]{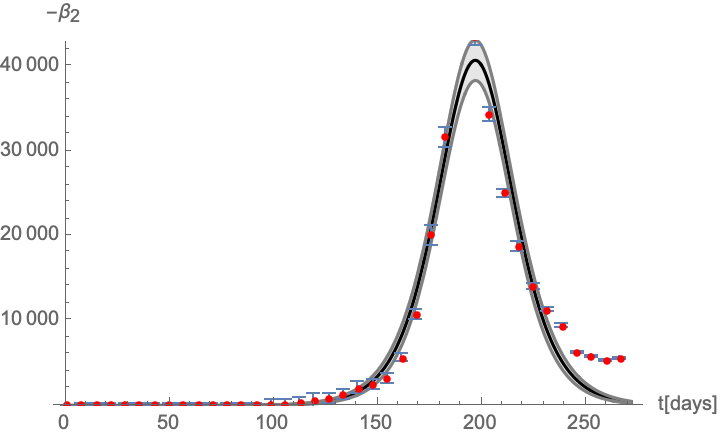}
\end{center}
\caption{Comparison of the time-dependence of the $\beta$-functions (\ref{GeneralBetaFunctionGradient}) for $(\Icn{1},\Icn{2})$ with the numbers of daily new infections in the UK.}
\label{Fig:UKQualTime}
\end{figure}

\begin{figure}[htbp]
\begin{center}
\includegraphics[width=7.5cm]{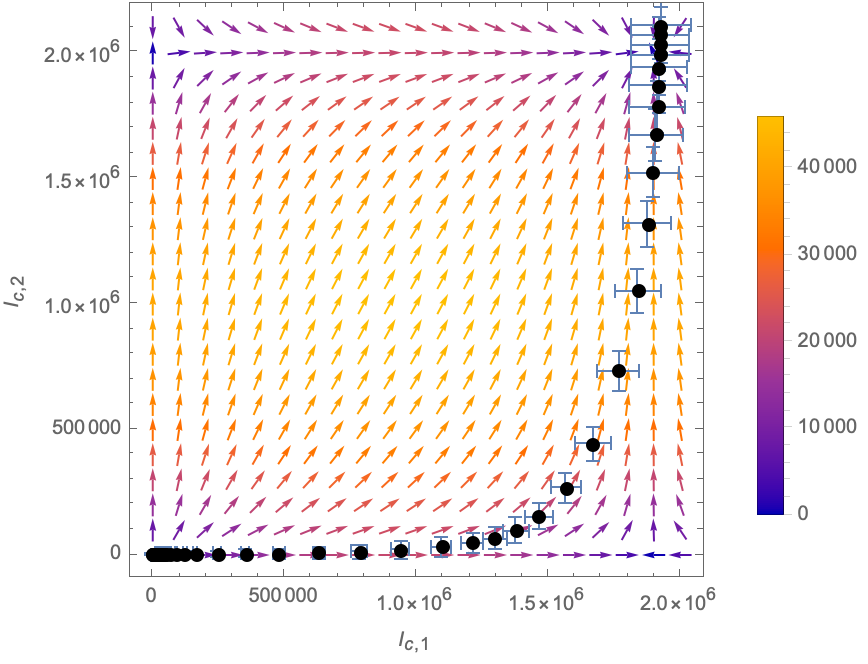}\hspace{1cm}\includegraphics[width=7.5cm]{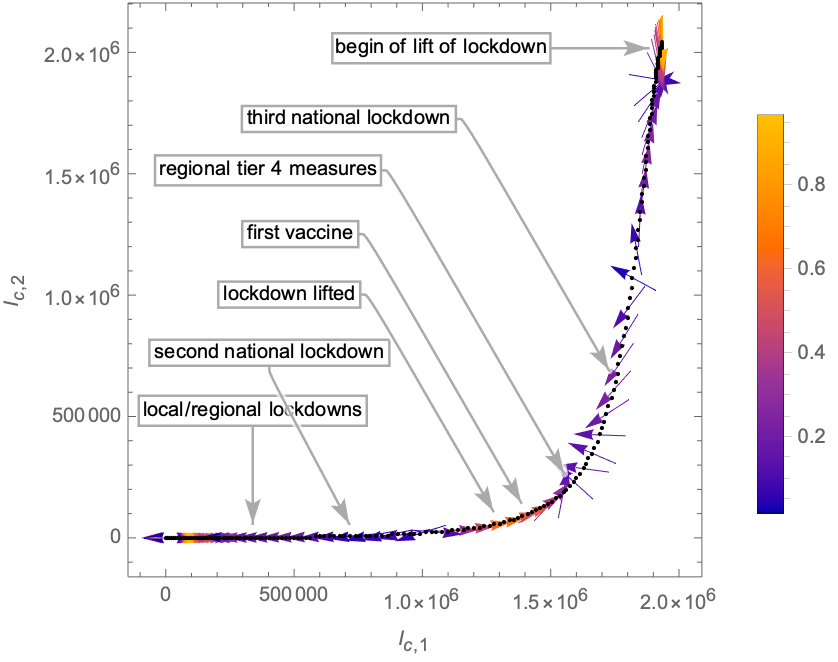}
\end{center}
\caption{Left panel: Vector field and flow of the system in the $(\Icn{1},\Icn{2})$-plane. Right panel: relative difference of the modelled $\beta$-function (\ref{GeneralBetaFunctionGradient}) with the actual number of daily infected along the flow of the system in the $(\Icn{1},\Icn{2})$-plane for the UK. As mentioned before, the deviations from the beta-function predicted by the eRG approach may be related to the changes in the lockdown measures or may indicate of additional subleading effects not captured in the current approach. }
\label{Fig:UKFlow}
\end{figure}

\begin{align}
&NA_1=(1.895\pm 0.035)\cdot 10^6\,,&&\kappa_1=136.06\pm 0.78\,,&&\lambda_1=0.0447\pm 0.0009\,,\nonumber\\
&NA_2=(2.007\pm0.030)\cdot 10^{6}\,,&&\kappa_2=197.12\pm 0.33\,,&&\lambda_2=0.0812\pm0.0014\,.
\end{align}
As in the case of California, these numbers are in fact determined by fitting the derivative of $\Icn{1,2}$ (\emph{i.e.} the daily new cases), as indicated in Figure~\ref{Fig:UKQualTime}. While the fit of $\beta_2$ correctly captures a single peak, the fit of $\beta_1$ gives a single maximum rather than two. The appearance of the second maximum in the red curve in the left part of Figure~\ref{Fig:UKQualTime} (stemming from the maximum of the orange curve in Figure~\ref{Fig:UK} in the end of Dec/2020) cannot be explained with statistical uncertainties inherent in the way we have extracted the relative number of new infections with the Alpha variant and may have different reasons:
\begin{itemize}
\item[\emph{(i)}] change of the infection rate of all variants due to modified social behaviour and/or travelling habits related to the Christmas holidays
\item[\emph{(ii)}] local geographic effects not captured by the data
\item[\emph{(iii)}] appearance of additional (subdominant) variants of the virus
\item[\emph{(iv)}] non-trivial interaction of the virus variants that are not captured by the beta-functions (\ref{GeneralBetaFunctionGradient})
\end{itemize}
Since the effect is rather small, we shall continue with the approximations in Figure~\ref{Fig:UKQualTime} and leave the analysis of this effect (notably the possibility \emph{(iv)}) for future work. With these approximations we can compute the function $\pot$ in (\ref{PotentialBetaFunction}) which in turn determines the $\beta$-functions (\ref{GeneralBetaFunctionGradient}). The corresponding flow of the system in the $(\Icn{1},\Icn{2})$-plane is shown in the left panel of Figure~\ref{Fig:UKFlow}, where we have chosen the scheme $f_i(\Icn{i})=\Icn{i}$. The right panel of that figure gives the relative deviation of the $\beta$-functions from the time derivative of $(\Icn{1},\Icn{2})$ as given by the number of daily new cases.

\begin{figure}[htbp]
\begin{center}
\includegraphics[width=7.5cm]{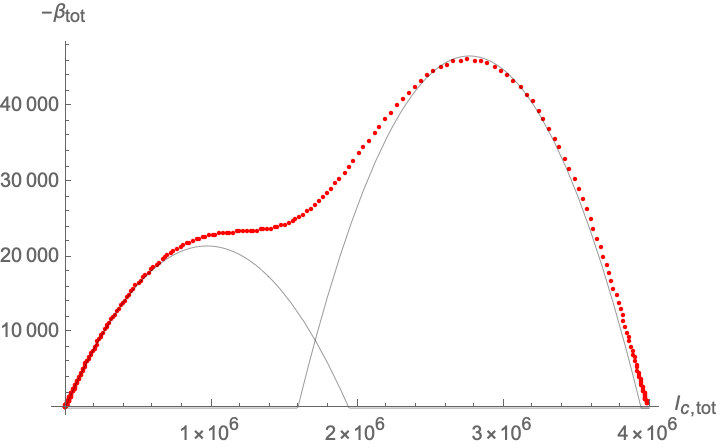}\hspace{1cm}\includegraphics[width=7.5cm]{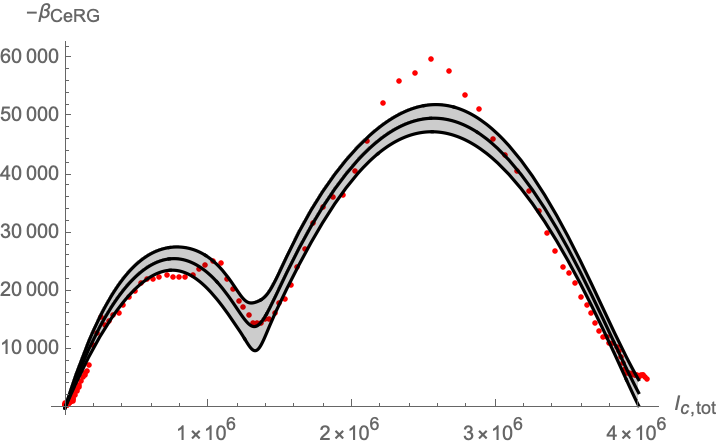}
\end{center}
\caption{Approximation of the time derivative of $\Icn{\text{tot}}=\Icn{1}=\Icn{2}$ (red dots). Left panel: $\beta_{\text{tot}}$ as defined in (\ref{CombinedTotalBeta}); Right panel: $-\beta_{\text{CeRG}}$ as in eq.~(\ref{CeRG}).}
\label{Fig:UKCrossOverBeta}
\end{figure}

From the flow diagram in Figure~\ref{Fig:UKFlow} we can see that the system remains close to the $\Icn{1}$-axis. We can therefore try to see whether it can be parametrised in a fashion resembling a crossover flow and whether it is possible to explain the two-wave structure from the flow close to a fixed point. To this end, we have plotted the time derivative of the total number of infected individuals $\Icn{\text{tot}}=\Icn{1}+\Icn{2}$ in Figure~\ref{Fig:UKCrossOverBeta} along with approximations along the lines of eq.~(\ref{CombinedTotalBeta}) and (\ref{CeRG}) (along with its $0.999$ confidence interval). Indeed, we can see that the function has a pronounced local minimum, which models the proximity of the system to the fixed point near the $\Icn{1}$-axis and which is responsible for the short linear growth phase in the end of November/beginning of December, as can be seen in Figure~\ref{Fig:UK}.

\subsection{First and second waves in South Africa}
\begin{figure}[htbp]
\begin{center}
\includegraphics[width=12.5cm]{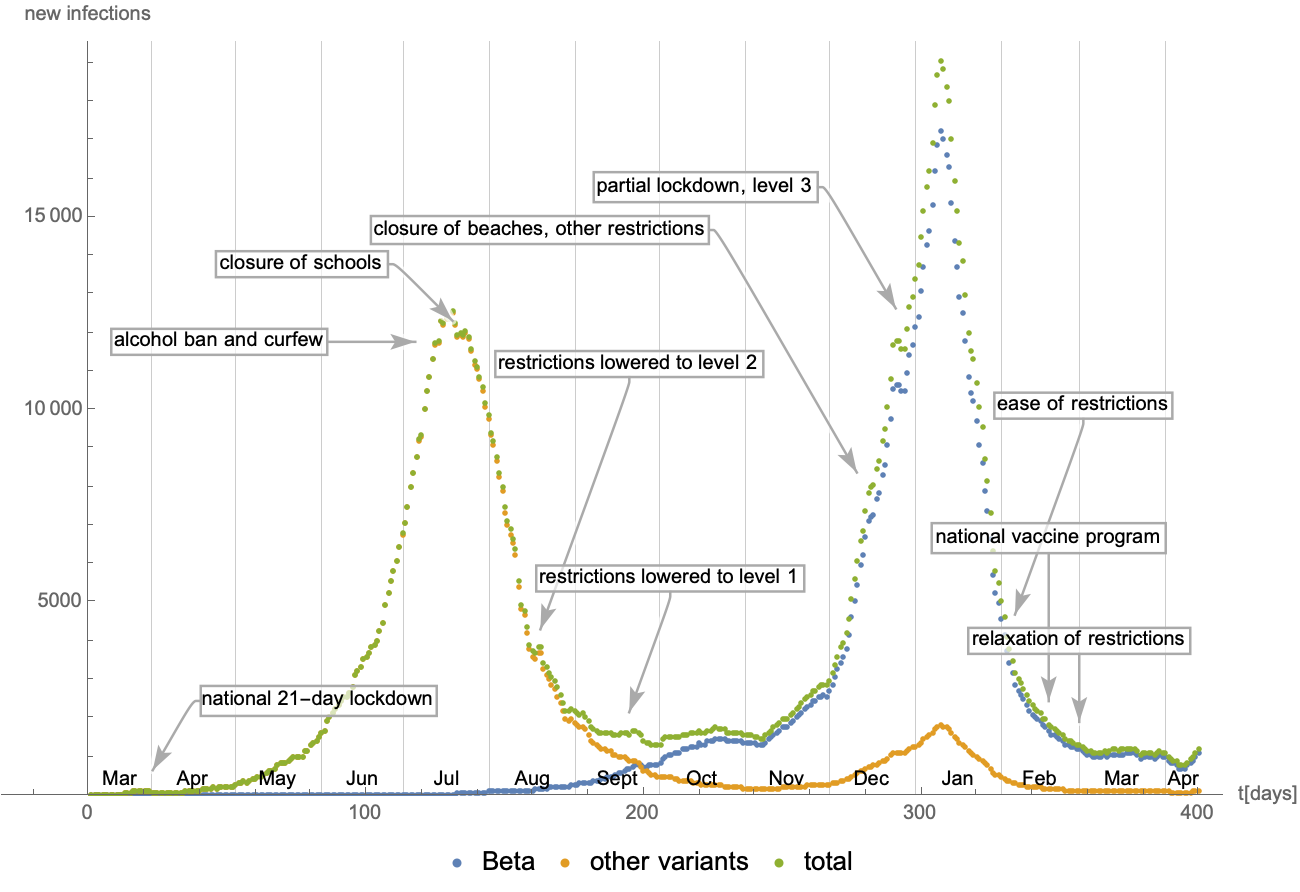}
\end{center}
\caption{Number of daily new infections in South Africa as a function of time since 08/March/2020. The epidemiological data have been extracted from \cite{OurWorldInData}, while the number of cases for the Beta variant are based on genome sequencing data extracted from \cite{Gisaid}. Finally, the information about lockdown and social distancing measures is taken from \cite{WikiSA}.}
\label{Fig:SA}
\end{figure}

\begin{figure}[htbp]
\begin{center}
\includegraphics[width=7.5cm]{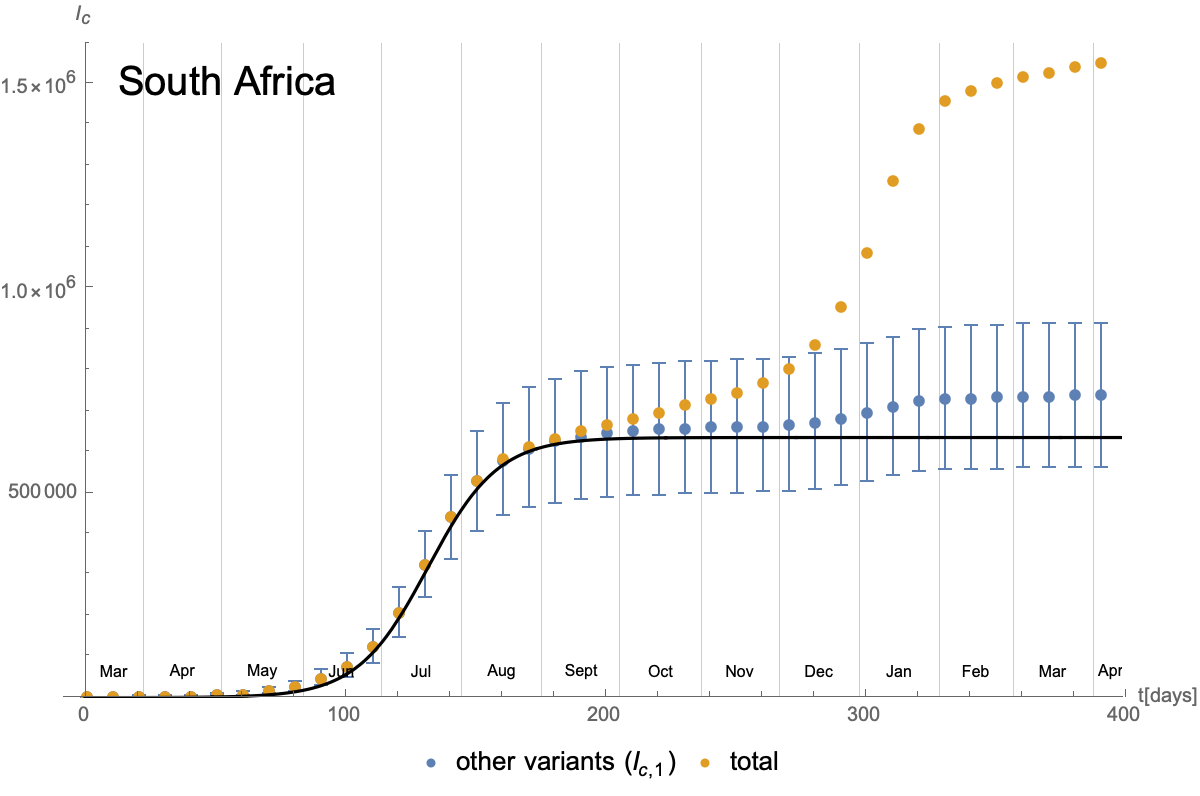}\hspace{1cm}\includegraphics[width=7.5cm]{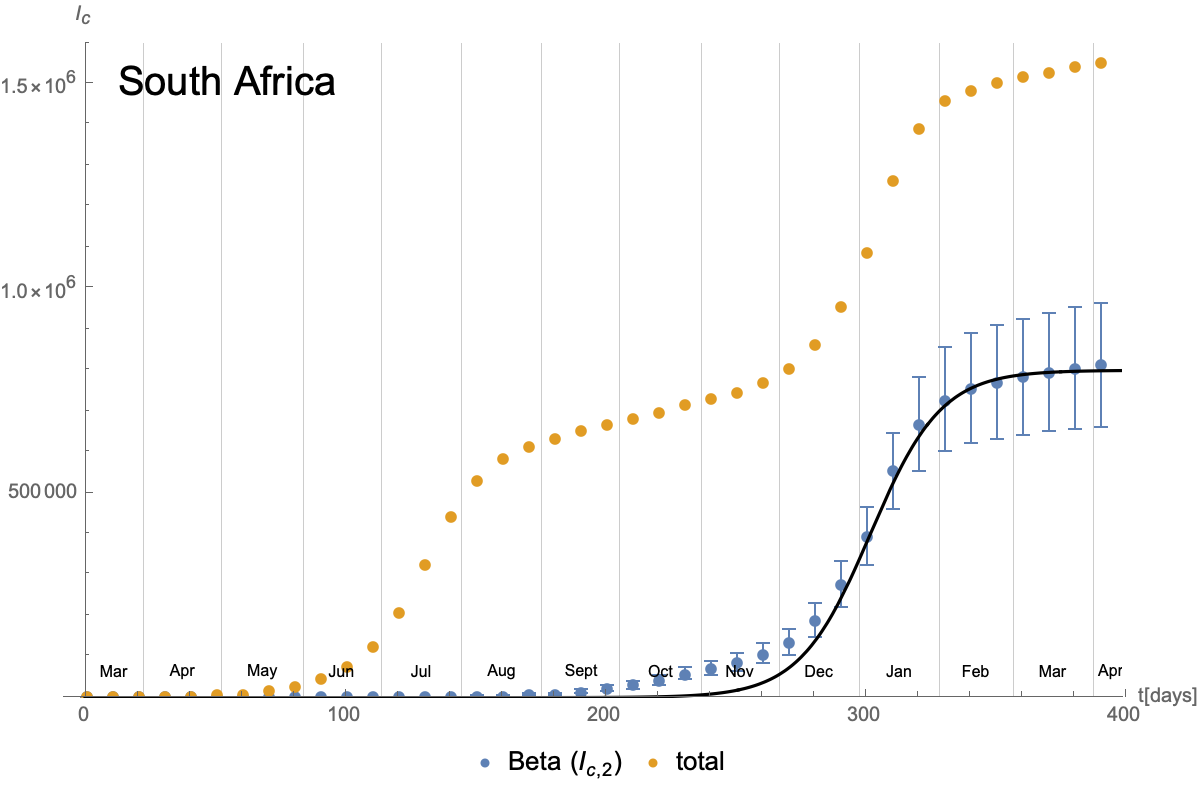}
\end{center}
\caption{Cumulative number of infected individuals for the variant B.1.351 (right panel) and all other variants (left panel) since 08/March/2020 and their approximation in terms of logistic functions. The error bars represent the (accumulated) statistical uncertainty following the rather low number of genome sequencings compared to the number of (positive) tests.}
\label{Fig:SAIc}
\end{figure}

As a final example, we consider the time evolution of the SARS-CoV-2 Beta\footnote{This variant is also called lineage B.1.351 under the pangolin tool and was first found in the Eastern Cape province of South Africa.} variant in South Africa. The daily number of new cases starting from 08/March/2020 is shown in Figure~\ref{Fig:SA}. The number of new infections with the Beta variant have been extracted using GISAID \cite{Gisaid} combined with epidemiological data from \cite{OurWorldInData}. Due to the rather small number of genome sequences of samples, the separation between the Beta variant and others is afflicted with a rather large (time-dependent) statistical uncertainty, which needs to be taken into account in the following and which makes the interpretation of certain results delicate. In Figure~\ref{Fig:SAIc} we have plotted the cumulative number of infected individuals alongside with their uncertainties. This figure also shows an approximation in terms of logistic functions, which is weighted by the (time-dependent) uncertainties: denoting the cumulative number of individuals infected with the variant B.1.351 by $\Icn{2}$ and the cumulative number of infected with the other invariants by $\Icn{1}$, the fit parameters along  eq.(\ref{LogisiticSol}) are given by

\begin{figure}[htbp]
\begin{center}
\includegraphics[width=7.5cm]{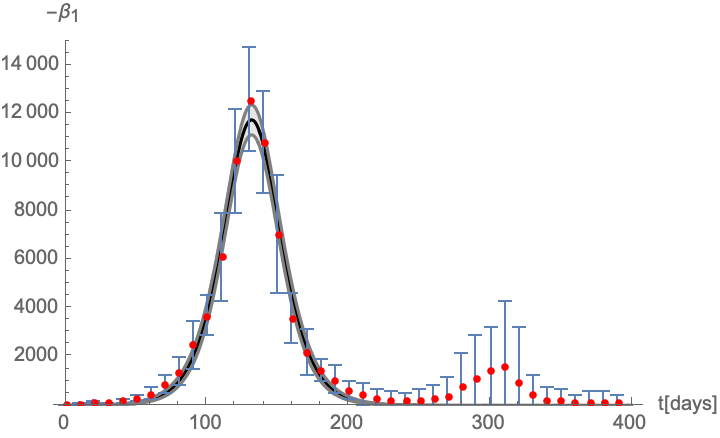}\hspace{1cm}\includegraphics[width=7.5cm]{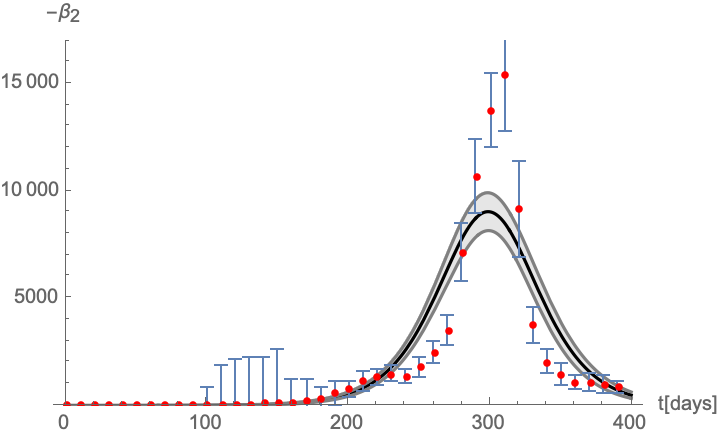}
\end{center}
\caption{Comparison of the time-dependence of the $\beta$-functions (\ref{GeneralBetaFunctionGradient}) for $(\Icn{1},\Icn{2})$ with the numbers of daily new infections in South Africa. The fits (along with their 0.99 confidence interval) are weighted with respect to the statistical uncertainty of the data. This explains for example, why the maximum of the red curve around day 300 in the left panel is effectively neglected: it is afflicted with a very high statistical uncertainty.}
\label{Fig:SAQualTime}
\end{figure}

\begin{figure}[htbp]
\begin{center}
\includegraphics[width=7.5cm]{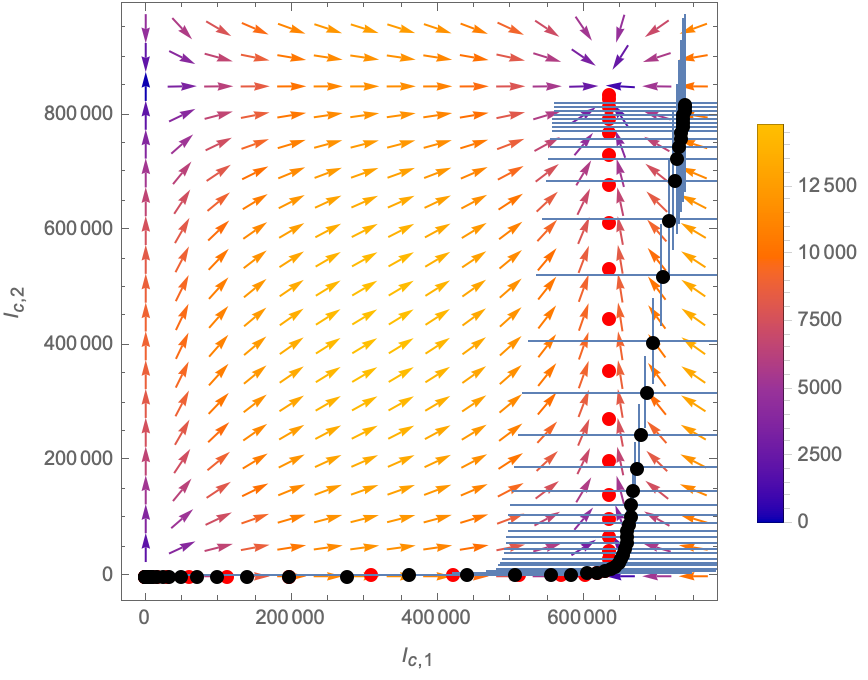}\hspace{1cm}\includegraphics[width=7.5cm]{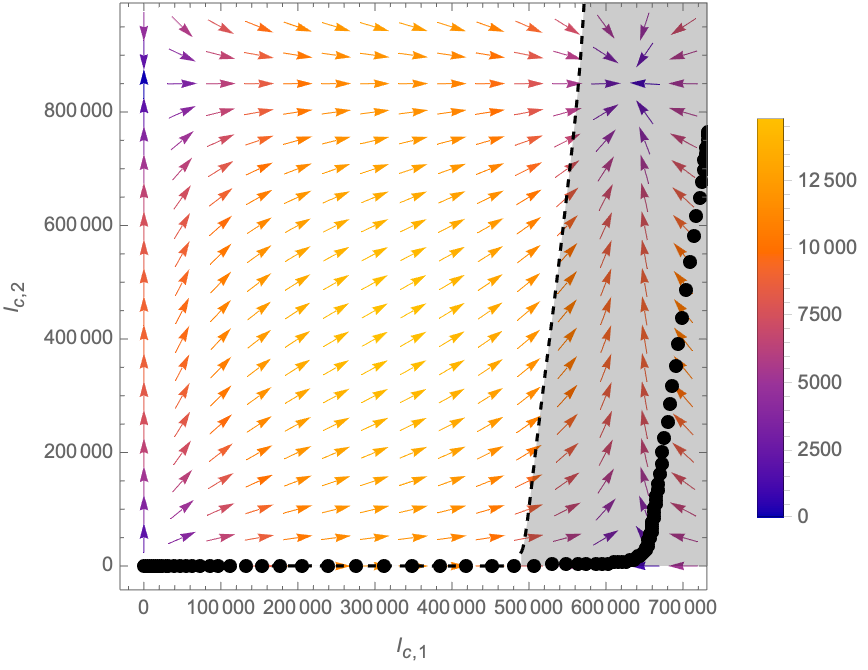}
\end{center}
\caption{Vector field and flow of the system in the $(\Icn{1},\Icn{2})$-plane. Left panel: the black dots represent the actual numbers of cumulative infected (together with their error bars), while the red dots represent the predicted eRG flow which lies within the error bars. right panel: For better visibility we have fused the error bars into the grey zone representing the statistical error of the cumulative numbers of infected individuals.}
\label{Fig:SAFlow}
\end{figure}

\begin{align}
&NA_1=(0.636\pm 0.009) \cdot 10^6\,,&&\kappa_1=131.86\pm 0.35\,,&&\lambda_1=0.0739\pm0.0012\,,\nonumber\\
&NA_2=(0.852\pm 0.020)\cdot 10^6\,,&&\kappa_2=298.91\pm 1.08\,,&&\lambda_2=0.0424\pm0.0007\,.
\end{align}
These numbers are determined by fitting (part of) the derivative of $\Icn{1,2}$ (\emph{i.e.} the daily new cases), as indicated in Figure~\ref{Fig:SAQualTime}. With these approximations we may compute the function $\pot$ in (\ref{PotentialBetaFunction}) which in turn gives an approximation of the $\beta$-functions (\ref{GeneralBetaFunctionGradient}). The flow of the system following the latter in the $(\Icn{1},\Icn{2})$-plane is shown in Figure~\ref{Fig:SAFlow} (as before, we use the scheme $f_i(\Icn{i})=\Icn{i}$). In the right panel, for better visibility, we have combined the error bars of the (black) data points into a gray region: while the black trajectory is not actually hitting the predicted fixed point of the beta-function, the latter is within the error bars. It is therefore difficult to say, whether this  indicates additional subleading effects in the time-evolution of the virus, or merely statistical uncertainty. Apart from this effect, the flow strongly resembles a crossover flow, which for the first part follows the $\Icn{1}$-axis, coming close to a fixed point at $(NA_1,0)$. However, the appearance of the variant B.1.351 triggers a crossover flow  to a new fixed point.

\begin{figure}[htbp]
\begin{center}
\includegraphics[width=7.5cm]{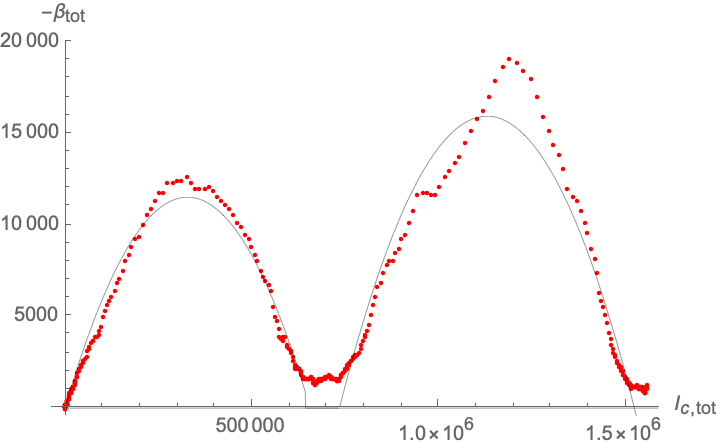}\hspace{1cm}\includegraphics[width=7.5cm]{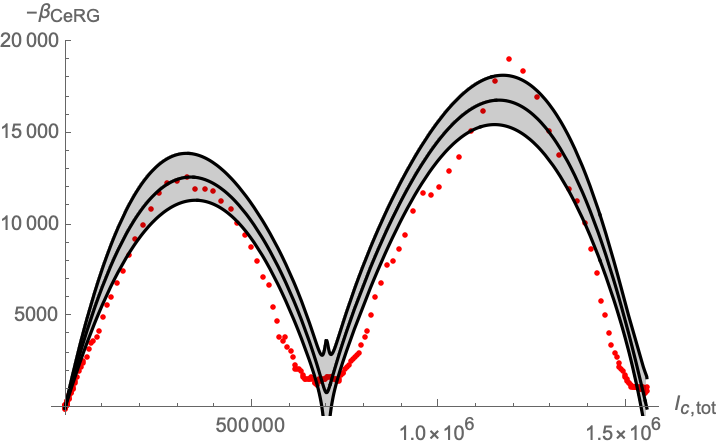}
\end{center}
\caption{Approximation of the time derivative of $\Icn{\text{tot}}=\Icn{1}=\Icn{2}$ (red dots). Left panel: $\beta_{\text{tot}}$ as defined in (\ref{CombinedTotalBeta}); Right panel: $-\beta_{\text{CeRG}}$ as in eq.~(\ref{CeRG}) together with its 0.99 confidence interval.}
\label{Fig:SACrossOverBeta}
\end{figure}

The fact that the flow in Figure~\ref{Fig:SAFlow} stays close to the $\Icn{1}$-axis during the first part of the flow, allows us to model it in the form of a crossover flow. To this end, we have plotted the time derivative of the total number of infected individuals $\Icn{\text{tot}}=\Icn{1}+\Icn{2}$ in Figure~\ref{Fig:SACrossOverBeta} along with approximations along the lines of eq.~(\ref{CombinedTotalBeta}) and (\ref{CeRG}). Indeed, we can see that the function has a pronounced local minimum, which models the proximity of the system to the fixed point near the $\Icn{1}$-axis and which is responsible for the linear growth phase between the end of August 2020 and the end of November 2020 as can be seen in Figure~\ref{Fig:SA}.

\section{Conclusions}
In this paper we have extended the epidemic Renormalisation Group (eRG) approach to study the time evolution of several competing variants of a disease. As a first step, in order to gain intuition, we have analysed a simple compartmental model (termed SIIR) with two different groups of infectious individuals (and different infection and recovery rates). Numerical solutions of the model indicate that as long as the reproduction numbers of the two variants are not too different, the cumulative number of infected individuals of the two variants can be well approximated by independent logistic functions (sigmoids). Moreover, we have approximated the dynamics in terms of flow equations that describe the trajectories of the system in the $\mathbb{P}$-plane keeping track of the cumulative number of infected. The resulting equations (\ref{BetaLin}) and (\ref{BetaQuad}) can be compactly formulated in terms of the gradient of a single function $\pot$, which is quadratic for variants with a reproduction number $\sigma<1$ and cubic for $\sigma>1$. Furthermore, from a theoretical perspective, we find it striking that the endpoints of the flows are characterised by co-dimension 1 surfaces in the $\mathbb{P}$-plane that represent equivalent system from the perspective of the SIIR model, akin to surfaces of fixed points that are related by the action of marginal operators in the context of conformal field theories.

As a second step, we have used the intuition gained from the simple SIIR model to propose a generalisation of the eRG framework to include the dynamics of multiple variants. Defining a beta-function (\ref{GeneralBetaFunctionGradient}) in the form of a gradient equation, we have analysed its fixed points in the $\mathbb{P}$-plane, as well as different trajectories connecting them. In particular, we have made contact to the CeRG approach \cite{cacciapaglia2020evidence,cacciapaglia2020multiwave} which has modelled the multi-wave structure of epidemics with the help of complex fixed points: in this regard, quasi-linear growth phases separating two waves are explained by the system coming close to a complex fixed point, which it cannot reach. In the current paper we have shown that such a behaviour can occur naturally through the appearance of a new variant of the disease.

Finally, we have confronted our model with data from the spread of different variants of SARS-CoV-2 in California, the United Kingdom and South Africa, thus empirically validating our approach.

In the future it will be interesting to further generalise and extend the approach developed here: on the one hand going beyond two competing variants will lead to a richer structure of fixed points for the system, thus allowing to model more complex multi-wave pandemics. On the other hand, so far our analysis has not taken into account other factors that govern the time evolution of pandemics, such as the impact of vaccines and non-pharmaceutical interventions but also the possibility for re-infections (through only partial immunity granted from recovery from a given variant). We hope to be able to return to these points in the future. Finally, we shall use the model developed in this paper for a computer-aided analysis of the spread of different SARS-CoV-2 variants in Europe and the USA along the lines of the work in our companion paper \cite{Companion}.


\bibliographystyle{ieeetr}
\bibliography{biblio}

\end{document}